# Scientific Promise

Enno Fischer

**Abstract**

Scientists constantly face decisions about what lines of research to pursue. This Element introduces the philosophical debate about scientific pursuitworthiness. It explains how it can be rational to pursue a theory even if the theory is less well supported than its rivals, and it discusses existing philosophical frameworks for guiding pursuit decisions. The Element also develops a new perspective. Existing accounts focus predominantly on theories, while experiments are largely neglected. This is an important shortcoming. Theoretical promise depends on experimental promise, and experimental promise raises questions of its own. Drawing on the epistemology of experimentation, the Element advances an account of experimental pursuitworthiness. It is argued that such pursuitworthiness depends on experimental virtues like a clear signal and simplicity of design. Moreover, the kinds of uncertainty that constrain the assessment of scientific pursuits are examined. Finally, the Element highlights open questions in the philosophy of scientific pursuitworthiness.

**Keywords:** pursuitworthiness, scientific promise, scientific rationality, theory choice, philosophy of experimentation

**Table of Contents**

# Chapter 1: From Promise to Pursuit

## 1.1 Two Challenges for the Methodology of Science

> How can we judge a theory's performance if we have not already made it a part of research? To object to the pursuit of an idea unless there is some guarantee in terms of performance is putting the cart before the horse, for the required guarantee can be obtained only by means of the very research one wants the guarantee for. (Feyerabend, 1981, p. 67)

What Paul Feyerabend describes here as "putting the cart before the horse" indicates an apparent dilemma. Whether a theory should be pursued or not depends on the future performance of that theory. That future performance, however, can only be evaluated when research on that theory is pursued. The decision to pursue or not to pursue a novel theory therefore appears to be a problematic one—no matter what course of action is taken. If we choose to pursue a novel theory, then this decision cannot be based on anything more than the *hope* that such efforts will bear fruit. If we choose not to pursue the novel theory, that will be a decision just based on the *superstition* that that theory was not worth the effort.

Talk about scientific promise matters because this is only an *apparent* dilemma. Decisions of pursuit can be based on more than 'hope' or 'superstition.' This Element is an introduction to the epistemological project of analyzing such scientific promise. It will review the guidance that philosophy of science has given on what scientific promise is, on how scientific promise can be determined, and how decisions on the pursuit of research projects should be made. The current Element will also advance a novel perspective on these matters. So far, discussions of scientific promise have largely focused on the promise of theories. The Element will show that philosophy of science is in a better position to provide guidance on scientific promise if it takes the promise held by experimentation into focus as well.

So, the first challenge here is that scientists need to make decisions under uncertainty. But in this regard scientists do not differ from investment bankers, poker players, or even anyone who travels on the notoriously unreliable German railway system. There is a second, deeper challenge here to our understanding of scientific inquiry as a rational endeavor, a challenge which is known as the *problem of innovation*: "if one insists [...] that standards for accepting a theory should be pretty demanding epistemically, then how can it ever be rational for scientists to utilize *new* theories which [...] will be likely to be less-well tested and well-articulated than their older and better-established rivals?" (R. Laudan & Laudan, 1989, p. 222f, emph. orig.).

To advance science, it seems scientists regularly need to make decisions that go against good reasoning. The problem poses a challenge to philosophers who want to understand science as a rational endeavor. If we aim to adequately *describe* scientific practices of theory development, then the problem of innovation appears to highlight strict limitations to seeing these practices as governed by good reasons. If we are looking for *normative guidance*, then the problem of innovation seems to put us in an awkward situation where we need to advise scientists to go against good reasons, or even to let go of such reasons entirely.

The two challenges for the methodology of science might seem avoidable if we can bracket pursuit decisions as non-rational. Accordingly, a traditional response to the challenges has been simply to admit that theory development is not governed by rationality and that not much of philosophical interest can be said about the challenges. According to Hans Reichenbach (1938) and other logical positivists, such challenges belong to the 'context of discovery.' This context, it is argued, should be subject to historical, sociological and psychological research that employs methods that are largely limited to describing—rather than evaluating—processes of theory development. Philosophers, by contrast, should focus on the 'context of justification,' which is concerned with normative or evaluative questions as to whether scientists are justified in their theory choices.

However, the issue of scientific promise sits ill with that distinction.[1] Scientific promise concerns the developmental stage of theory building. It is important even before discovery because it informs scientists about where to look for new insights. At the same time, it acts as a justification for taking up certain pursuits. Thus, a strict separation between discovery and justification is not possible. Neither is it theoretically desirable. Insofar as philosophy of science has a genuine interest in understanding scientific practice, it cannot and should not be limited to investigations of justification, and it cannot and should not sideline the issue of scientific promise.

That theory choice is a central subject in the philosophy of science is uncontroversial. But philosophers' talk of 'theory choice' is notoriously ambiguous. It can refer to what theories scientists choose to accept but also to what theories they choose to invest their efforts in. And often it is that latter choice what leads to disagreement among scientists. Consider a classic example: disagreement between proponents of Ptolemaic and Copernican research traditions in the sixteenth century (see, e.g., Kuhn 1977). Eli Lichtenstein (2021) has argued that this disagreement is most accurately described as one about scientific promise. When Copernicus decided to promote heliocentrism, he evidently did not make a choice between two complete frameworks. At that point heliocentrism simply hadn't been developed into a complete framework yet. Instead, it was Copernicus's judgement that, first, "Ptolemaic systems were unlikely to become both simpler and more accurate via continued internal revision" and, second, "that he might do better by means of a more fundamental conceptual reorientation" (Lichtenstein, 2021, p. 169). What mattered for Copernicus's disagreement with his contemporaries was not primarily what framework was to be accepted but what framework could be expected to be better if further efforts were invested.

[1] Reichenbach's dichotomy between discovery and justification has long been challenged also for many other reasons (Hoyningen-Huene, 1987; Schickore & Steinle, 2006).

Note that the problem of innovation is a threat to *scientific* rationality. What scientific rationality amounts to is, of course, a question that will defy simple answers. Yet, it certainly is a kind of epistemic rationality, insofar as science has the goal to improve and increase knowledge of the world. There are other kinds of rationality that play a role in scientific practice that are not or need not be primarily of that epistemic kind. These are particularly evident in situations where rational credit-maximizing behavior misaligns with the goals of science. For example, scientists are incentivized "to produce more results at the expense of spending more time on the reproducibility of any given result" (Heesen, 2018). Such "rushing into print" (ibid.) may well be perceived as rational from a career-strategic perspective, but as this kind of behavior does not (or at least not necessarily) promote the epistemic goals of science, it is not the kind of epistemic rationality that would need to be recovered in order to address the problem of innovation.

Philosophical reasoning about scientific promise, thus, has two related but differing goals. Both will be discussed in this Element. First, such reasoning is motivated by the practical question for guidance in selecting the most promising lines of research. Insofar as 'scientific promise' is a subject of general philosophy of science or methodology of science, that practical question is unlikely to be answered conclusively when it comes to concrete decision making. But a philosophical account of scientific promise can be informative without necessitating concrete decisions, when it explicates the values that do and should play a role in the assessment of scientific promise. This Element will contribute here by explicating values that matter for experimentation.

Second, there are epistemological questions regarding the rationality of scientific pursuits. Can science be understood as governed by good epistemic reasons? What kind of reasons are they? Such questions of rationality, in turn, have a deep significance for the justification of science's role in our society. The belief in scientific rationality is the reason many contemporary societies are ready to spend large amounts of public funds on science. This Element will contribute to the project of understanding scientific rationality by advocating for a more comprehensive approach to scientific promise that also includes experimentation.

**1.2 Pursuitworthiness**

A promise is an assurance that one person can give to another. Relatedly, individuals or things can be promising or hold promise. A promising student may not be able to answer all questions on the exam yet but has qualities such as talent and diligence that make us believe they will be able to do so in the future. A promising writing proposal may not provide details of the final manuscript but may still give us reasons to expect that the final manuscript will be of good quality if further developed.

*Scientific* promise concerns the future success of research. While a theory may not yet have all the qualities that would make it a good theory, we may have reasons to expect that this will be the case if we put more work into it. While it may be unclear what the exact result of an experiment will be, we may have expectations that the outcomes will constitute a significant advancement for science. Importantly, scientific promise is of a conditional form: it is a promise that will only be fulfilled if efforts are made by the scientists. Before they can reap any of the gains, scientists need to develop theories and models, they need to set up research facilities and conduct experiments. Scientific promise is what motivates scientists to pursue a project. It is what makes a project worthy of pursuit, or for short, *pursuitworthy*.

The concept of pursuitworthiness has been at the center of philosophical discussions of scientific promise and it will be the key concept discussed in what follows. There are a few other concepts that are related to issues of scientific promise and that have been discussed in the literature to be covered here. For instance, Kuhn's (1977) notion of *fruitfulness* as a criterion—or 'value'—for theory choice has received some recent discussion (Haufe, 2024; Ivani, 2018). Other concepts are fertility (McMullin, 1976; Nolan, 1999; Schindler, 2017), heuristic appraisal (Nickles, 1989, 2006), promisingness (Shan, 2020), and creativity (Sánchez-Dorado, 2023). I opt for the somewhat technical notion of pursuitworthiness because I am interested in warranted decisions to put one's efforts into researching a theory. A fruitful, fertile or creative theory may not be pursuitworthy, simply because there are other theories that are more fruitful, fertile or creative. Likewise, a theory may be pursuitworthy even if it is not very fruitful, simply because it is still the best way of investing one's efforts.

Before we dive deeper into discussions of pursuitworthiness that concept should be disambiguated in three regards. First, we distinguish between *ex-ante* and *post-hoc* pursuitworthiness. The ex-ante pursuitworthiness of a research project concerns the expected value of that research project before it is pursued and assesses scientific research in a forward-looking mode. Post-hoc pursuitworthiness, by contrast, concerns evaluations made after a project has been pursued, once its full value is evident. Post-hoc pursuitworthiness is backward looking in that it looks at past successes and failures. Most discussions of pursuitworthiness concern the ex-ante perspective: this is the kind of judgement that is required for decision making, and it is relevant for understanding the rationality of decisions. Post-hoc assessments of pursuitworthiness can also play an important role, for example, when we look at a theory's track record to inform our expectations regarding its future performance. Finding out that pursuing a project was not worth the effort may give important information for future decisions regarding that project. If, by contrast, a project has a strong track record, that may count in its favor when it comes to extending research efforts directed at it. Post-hoc assessments of pursuitworthiness, thus, can be an important resource for evaluating ex-ante pursuitworthiness.[2]

The focus of this Element will be assessments of ex-ante pursuitworthiness. Such assessments can be made at various stages of inquiry. Most common are discussions of nascent theories, when scientists are confronted with the decision to invest their efforts into ideas that are not yet able to compete with established ideas but may show the promise to do so in the future. But considerations of ex-ante pursuitworthiness are also raised regarding developed theories that have already been pursued for an extended time. Will it pay off to invest further resources into such a theory in addition to those resources that have already been invested? Or should research on this theory be abandoned?

Second, we distinguish between *comparative* and *absolute* notions of pursuitworthiness. Comparative pursuitworthiness concerns whether one project is more pursuitworthy than another. This notion is important when there are several options for further pursuit and it needs to be decided which option or set of options are to be prioritized over the others. Major difficulties that arise here concern the comparability of the outputs of competing projects. By what criteria can we show that the output of one project is (likely) more valuable than that of other projects? Absolute pursuitworthiness concerns whether a project is pursuitworthy at all. Will the research effort pay off sufficiently to justify the efforts that have been invested? Here a major challenge lies in establishing a comparison between the efforts invested and likely outcomes to be gained. For example, is there a systematic way to relate epistemic outputs to research funds spent, especially when the epistemic outputs are intended to advance foundational research?

Third, we distinguish between *epistemic* and *non-epistemic* pursuitworthiness. Epistemic pursuitworthiness concerns the value of pursuing a project for the purposes of increasing and improving knowledge. Knowledge can be increased and improved, e.g., by developing and testing new theories and models, by collecting new data, generating novel hypotheses, etc. Non-epistemic pursuitworthiness concerns the value of a project that is not concerned with improving and increasing knowledge, but concerns, for example, the economic benefits of pursuing a scientific project or the capacity of a project to improve life quality. The distinction between epistemic and non-epistemic pursuitworthiness is not sharp. Whether and to what degree a scientific project increases and improves knowledge in relevant ways depends on epistemic as well as non-epistemic factors. For example, a project of basic research may be pursuitworthy because it addresses many important theoretical questions that need to be addressed such that a research field such as cancer research may advance. Ultimately, however, the advancement of that field will still be motivated by non-epistemic goals, for example, curing cancer.

[2] Note that post-hoc assessments of pursuitworthiness come with the risk of anachronism if past efforts are evaluated on present standards. See Furlan (2022) for additional reflection on this point.

The focus of this Element will be *ex-ante* evaluations of *epistemic* pursuitworthiness of both absolute and comparative kinds. Post-hoc evaluations of pursuitworthiness will be addressed insofar as they support evaluations of ex-ante pursuitworthiness. Moreover, non-epistemic notions of scientific pursuitworthiness are an exciting topic for philosophical debate, including urgent questions such as how science best facilitates the needs of democratic societies (Intemann, 2015; Kitcher, 2011). However, in this Element such issues will be mostly sidelined and only discussed insofar as a clear separation of epistemic and non-epistemic goals of scientific endeavors cannot always be made.

### 1.3 The Pursuitworthiness of What?

Another issue that needs to be clarified right away is what the ‘unit of appraisal’ is. What is being pursued? And the pursuitworthiness of what is to be evaluated or appraised? One can find various proposed units of appraisal, including research programs (Lakatos, 1978), research traditions (L. Laudan, 1977), theories (L. Laudan, 1977; e.g., McMullin, 1976; Whitt, 1990), models (Han, 2023; Haueis & Kästner, 2022), cognitive systems (Šešelja & Straßer, 2014), ideas (Duerr & Fischer, 2025), projects (Shaw, 2022), questions (Barseghyan, 2022; DiMarco & Khalifa, 2022; Wilholt, 2020), and experiments (Fischer, 2026b; Laymon & Franklin, 2022). Of these, theories and the theoretical parts of research programs have by far received the most attention.

A main lesson of this Element is that an adequate picture of scientific promise can only be given if we go beyond the narrow focus on the pursuitworthiness of theories and include discussions of experimental pursuitworthiness. An overly narrow focus on theories is problematic for two reasons. First, theoretical pursuitworthiness depends on experimental pursuitworthiness. We will see that there is an overarching consensus that theory promise to a large degree depends on the theory's potential to address empirical problems. This potential, in turn, can only be unlocked if there is an appropriate empirical methodology that comes with the theory. Second, experimental pursuitworthiness raises exciting questions of its own. Most importantly, shifting the focus to also include experimental pursuitworthiness will suggest a new set of pursuitworthiness guidelines and a shift in the prioritization of extant guidelines. More precisely, the Element will survey a series of experimental virtues such as precision, broadness of sensitivity and experimental simplicity and explain how these virtues do and should influence considerations of pursuitworthiness.

This Element limits itself to a general epistemological discussion of pursuitworthiness without any deep dives into case studies. Examples (mainly from physics) will only be employed occasionally to illustrate the overarching points. As a consequence, the Element will not cover the more local and discipline-specific factors that affect the pursuitworthiness of concrete research projects. Moreover, the current introduction is limited to philosophical research on pursuitworthiness and will not cover insights from economics of science, social studies of science, and other neighboring disciplines.

### 1.4 Overview

Chapter 2 discusses the pursuitworthiness of theories, which is the focus of existing philosophical debates. It will first survey approaches to delineate two distinct modes of assessing theories, one geared towards the acceptance of theories and the other geared towards theory pursuit. Chapter 2 will then discuss a series of indices of theory promise that have been proposed in the philosophical literature. Towards the end of chapter 2 I will turn to more recent approaches that address issues of decision making: (how) can indices of theory promise be put to work such that well-informed choices can be made? Chapter 3 will argue for a more comprehensive picture of scientific pursuitworthiness that incorporates experimentation. While philosophers of science have made great advancements in the epistemology of experimentation, these advancements have been recognized in the literature on pursuitworthiness only to a very limited degree. Chapter 3 will begin by explaining why a closer look at experimentation is essential for an adequate picture of scientific promise. It will then go on to review some key insights from the philosophy of experimentation that merit consideration and explicate what is meant by identifying an experiment as pursuitworthy. Chapter 3 will then turn to a more systematic discussion of what makes an experiment pursuitworthy. More specifically, it will identify experimental virtues that do and should play a role in the assessment of experimental pursuits, and it will discuss the role of various kinds of uncertainty in such assessments. Chapter 4 wraps up the discussion and provides a brief outlook on future research topics in the epistemology of pursuitworthiness.

## Chapter 2: Pursuitworthy Theories

### 2.1 Acceptance and Pursuit

It seems advisable to base one's research only on theories that are well-established. However, if all scientists focus their work exclusively on such theories, how can there ever be new theories? It seems that to be able to compete with established theories any newcomer theory would need to get a chance to accrue initial support. But that is possible only if at least some scientists go against the advice of basing their research only on well-established theories. This is what I have identified in chapter 1 as the problem of innovation, a central challenge for understanding how science progresses.

The initial motivation of introducing concepts of pursuitworthiness was to tackle this epistemological problem. More specifically, the concept of pursuitworthiness can be traced back to a distinction between two contexts or modes of appraisal, introduced by Larry Laudan (1977)[3]: the context of acceptance and the context of pursuit.[4] In the context of acceptance scientists are concerned with selecting "among a group of competing theories and research traditions" the one that is to be treated "as if it were true" (1977, 108). In the context of pursuit, by contrast, scientists decide which theories and research traditions to work on, investigate, or explore. According to Laudan, these are often theories and research traditions that are "patently less acceptable, less worthy of belief, than their rivals" (110). The problem of innovation can be solved, it is suggested, by showing that decisions to pursue novel theories are governed by criteria that differ from those governing theory acceptance. To see what this means, let us specify in more detail the difference between the two modes of appraisal.

Accepting a theory comes with strong commitments to believe in the theory and its empirical consequences. It involves "the belief that the theory provides the best of all available explanations; the belief that it is empirically adequate, or the most effective problem-solver in some domain; and if one is a realist, the belief that the theory is (approximately) true and that the theoretical entities posited by it exist" (Whitt 1990, 471). Pursuing a theory, by contrast, does not necessarily come with a strong belief in the theory and its empirical consequences. We may pursue a theory even if it is not thought to be empirically adequate, an effective problem-solver, or (approximately) true in its current form.

[3] Pursuitworthiness and scientific decision making have, of course, been discussed by philosophers of science well before the late 1970s, especially with regard to science 'planning' (for an overview see Baker (2022)). The discussion here will focus on developments since Laudan's introduction of the concept.
[4] Elsewhere Laudan speaks of "a whole spectrum of cognitive stances that scientists can adopt toward their theories (suggested by phrases like 'entertain', 'consider', and 'utilize as a working hypothesis') (L. Laudan, 1996, p. 111).

Instead, pursuing a theory means to "work on" that theory. This can mean that one is primarily concerned with improving the theory itself. Such pursuits aim at establishing the theory as a genuine alternative to existing theories (Laudan 1977). Think, for example, of early theoretical developments in quantum mechanics that helped establish that theory. Alternatively, to work on a theory can mean "to engage" it "or to make use of it in some portion of one's research activities" (Whitt 1990, 470). Such pursuits employ a theory to promote other theories or parts of a discipline or neighboring disciplines that are not primarily concerned with establishing or supporting the theory that is being employed. For example, think of applications of quantum mechanics that have helped develop chemical theories.

Thus, accepting a theory is justified by an assumed superior epistemic status of that theory, whereas pursuing a theory is justified by the novel insights that may result from working on it, that is, its heuristic value. What is the relation between epistemic status and heuristic value? Ernan McMullin (1976) argues that there are important ways in which the two can come apart:

> To say that a theory has a high degree of heuristic potential tells us nothing of its epistemic status. Indeed, the greater its heuristic potential, the more problematic its present epistemic status may well be, since it contains so much in the way of untested suggestion. Likewise, to speak of a theory as 'established' tells us nothing of its heuristic potential for the future. Some of the best-established theories have very little research potential [...]. (424)

McMullin makes two points here. First, if the heuristic potential of a theory is high, that does not imply that it has a high epistemic status. Consequently, high pursuitworthiness does not imply acceptability. For example, Barry Marshall and Robert Warren's early theory that peptic ulcers are caused by bacteria had considerable heuristic potential. Yet, it was not acceptable at the time because much current research favored the competing acid theory (Fleisher 2022). McMullin also indicates that one can go even a step further, saying that it is exactly the shortcomings of a theory that can be indicators of their heuristic potential. In a similar vein, Lichtenstein (2021) argues that scientists "can be attracted to a theory precisely by virtue of its explanatory weaknesses" (170), mentioning the inconsistencies within early Copernican astronomy as an example. An explanatory weakness by itself, of course, does not justify pursuit of a theory but only against the backdrop of an otherwise promising approach (ibid.).

Second, if a theory has secured a certain epistemic status that does not imply that it has heuristic value. Consequently, acceptability does not imply pursuitworthiness. Thomas Nickles (1989) illustrates this point by contrasting Hilbert's contributions to the theory of invariants with Watson and Crick's discovery of the molecular structure of DNA. Both achievements represent solutions to long-standing scientific problems. But Hilbert's theory of invariants marked the end point of a research program, whereas Watson and Crick's achievement started a whole new research field by suggesting a copying mechanism for genetic material.

Note that Laudan's talk of contexts of pursuit and acceptance implies a temporal order: first pursuit then acceptance.[5] This order matters for addressing the problem of innovation. We need to explain why scientists research theories *before* they are acceptable. The temporal order, however, does not hold in general. As indicated by Nickles's examples, questions of pursuitworthiness also concern theories that are accepted.

What follows from this preliminary characterization of acceptance and pursuit? First, we can distinguish between *communal* and *individual* pursuitworthiness (Šešelja et al., 2012). What is pursuitworthy for a research community as a whole may not be pursuitworthy for an individual with concrete training, skills, research experience, and access to facilities. That distinction does not apply as straightforwardly to acceptance. If an individual accepts other theories than the research community, that mismatch calls for an explanation.

Second, we can distinguish between claims of pursuitworthiness that come in the form of *research directives* and claims that come in the form of *evaluative stances* (Šešelja et al., 2012). A research directive indicates what theories an individual scientist should pursue, given their specific research context. An evaluative stance, by contrast, is an answer to the more general question of what pursuits would be in the epistemic interest of a scientific discipline or science as a whole. The difference is important because it allows one to appreciate pursuits even if one cannot contribute to them. For example, I can have the evaluative stance that empirical research on climate change is highly pursuitworthy. But there is no corresponding research directive that would force me as a philosopher to pursue such research because it requires an entirely different skill set.

Third, the context of pursuit is sometimes characterized as being more compatible with scientific pluralism. Accepting mutually excluding theories can lead to considerable cognitive dissonance. But the same does not apply to pursuit: "a scientist can often be working alternately in two different, and even mutually inconsistent, research traditions" (Laudan 1977, 110, emph. orig.). Moreover, scientists may follow different research directives while they have the same evaluative stances. Thus, scientists pursuing one theory or research tradition do not have to disagree about pursuitworthiness with scientists pursuing other theories or research traditions (Šešelja et al., 2012; Šešelja & Straßer, 2013).

[5] A similar temporal order is suggested by Curd (1980).

Finally, there is sometimes a tendency to describe the criteria for pursuit as being weaker than those for theory acceptance (see, e.g., Achinstein (1993)). In a sense this is right: as we will go on, we will see that analogies are commonly thought of as playing a vital role in justifying theory pursuit but considered to be too weak as a reason for theory acceptance. But to show why it can be rational to focus one's research on non-acceptable theories, we need to show that the context of pursuit is governed not just by a watered-down version of the criteria for acceptance. We need to identify a distinct set of criteria, criteria according to which non-acceptable pursuitworthy theories can score higher than acceptable ones. In what follows, I will first discuss concrete examples of such indices (section 2.2), then examine their epistemic status (section 2.3), and finally address whether and how they can guide decisions (sections 2.4 to 2.6).

### 2.2 Indices of Theory Promise

According to Laudan, what matters for acceptance is whether a theory represents *progress*. A progressive theory is one that solves a larger total number of significant problems (both empirical and conceptual) than all competing theories. What matters for pursuit, by contrast, is the *rate of progress*. The rate of progress is the number of new problems solved per time unit. Laudan argues that "it is always rational to pursue any research tradition which has a higher rate of progress than its rivals" (1977, 111).

This distinction allows us to identify theories that are more pursuitworthy than their competitors, even if they are less acceptable. An example is Dalton's atomism (Laudan 1977, 113). The dominant research tradition in the chemistry of Dalton's time was the theory of elective affinities, which sought to account for chemical reactions by positing that chemical elements prefer to bond with some elements rather than others. At the time Dalton proposed his atomistic theory, elective affinities theory had accrued support for a century and could explain many chemical phenomena. Moreover, Dalton's atomism was confronted with serious problems (Whitt, 1988). However, it predicted that chemical substances combine in certain definite ratios and multiples thereof. This gave rise to a lively research program of determining relative atomic weights from the relative weights the elements have in chemical compounds (see Chalmers & Coko, 2026 for details).

Laudan's approach has some initial plausibility as an explanation for why it can be rational to pursue a theory that is not as well-established as its strongest competitors. If a research field develops fast, it tends to also generate new questions and problems at a fast rate. Consequently, one can hope to pick the low-hanging fruit by joining these research efforts. Conversely, if a long time has passed since a research tradition has seen significant advancements, then it is likely that those low-hanging fruits have all been picked already.

However, there are also several problems with Laudan's rate-of-progress criterion. First, the criterion assumes that there is a track record that can be the basis of evaluating a theory's rate of progress. But insofar as pursuit is concerned with nascent theories there is no track record. A high rate of progress is only possible if there are already scientists in place who deem that theory pursuitworthy. On what rational basis did those first scientists decide to pursue the new theory? It seems that Laudan's criterion doesn't really solve the problem of innovation if it does not explain that initial motivation.

Second, Laudan's criterion of pursuitworthiness rests on an implicit assumption that past and present trends in the rate of progress are extrapolatable to the future. In Laudan's account, the rate of progress is derived from the recent and current rate of problem solving. This may work as a criterion for post-hoc pursuitworthiness. For this to count as a criterion also for ex-ante pursuitworthiness one needs to assume that the past rate of success represents a trend that is sustained at least over the immediate future. But this need not be the case. Whether a certain trend continues if additional time is spent on the project or not is exactly the question that is at stake in assessments of ex-ante pursuitworthiness.

To see the challenges of extrapolation, consider the case of high-energy physicists searching for new fundamental particles. After the famous discovery of the Higgs boson in 2012, physicists have been searching for indicators of new particles for more than a decade without any conclusive findings to date. Thus, the corresponding research program can be described as having a low rate of progress. When physicists are now advocating for new multi-billion-dollar particle colliders, they must explain this recent weak track record. However, that weak track record is certainly not sufficient to deny proposals of new colliders, because it is unclear whether that trend will continue once physicists have new machines available to probe higher energy regimes.

An additional issue that arises here is that of choosing the right time interval for evaluating a theory's rate of progress. Focusing on the recent past one may get the impression that there was relatively little empirical advancement in particle physics. Extending the time interval to include the successes of the twentieth and early twenty-first centuries, however, would give the impression of a highly successful research program. Since Laudan does not specify how the time interval is to be chosen, his criterion does not give a clear verdict on this case.[6]

A clearer distinction between ex-ante and post-hoc modes of evaluation is drawn by McMullin (1976). McMullin distinguishes P-fertility (proven fertility) and U-fertility (untested fertility). P-fertility is relevant for theory acceptance. It relates to a theory's "performance, not potential; it is estimated by the *actual* success of the theory" (400). To evaluate this performance, one needs to look into the theory's past development and how the theory has dealt with anomalies and problems. U-fertility, by contrast, is related to the theory's future research potential. A theory's U-fertility depends on the theory's likelihood of giving rise to interesting extensions, solutions to outstanding problems, to "unify hitherto diverse areas" and to "open up entirely new territory" (423f).

McMullin's notion of U-fertility helps us to put a finger on what is missing in Laudan's approach: an assessment of a theory's capacity to give rise to new research. However, McMullin's approach leaves unspecified what that capacity consists in, and how we can track it. Moreover, McMullin draws an overly sharp line between assessments of past performance and future potential. It seems plausible that a theory's past performance can act as a useful heuristic for assessing its future potential.

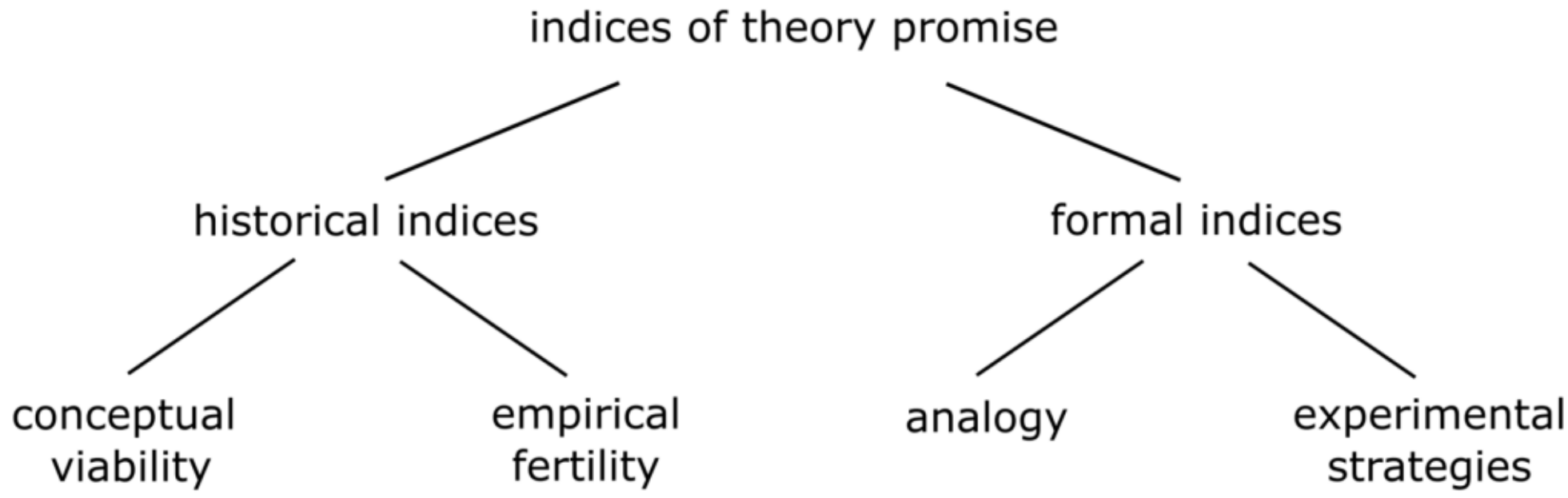

*Figure 1: Whitt's (1992) taxonomy of indices of theory promise. Historical indices reflect a theory's past performance. Formal indices are current characteristics of a theory that make it pursuitworthy.*

Laurie Anne Whitt (1992) makes considerable progress by synthesizing both Laudan's and McMullin's criteria. Whitt's taxonomy (figure 1) distinguishes between historical and formal indices. *Historical indices,* like Laudan's rate-of-progress criterion, take a theory's performance over some past time interval into account. There are two such kinds of indices in Whitt's model. *Conceptual viability* is a theory's demonstrated ability to "undergo conceptual growth and refinement as the result of its deployment in research" (617). Key factors are that relevant theoretical concepts can be refined, that conceptual resources from theories in other domains can be integrated, and consilience, the theory's ability to account for a large variety of independent phenomena. *Empirical fertility* concerns a theory's "demonstrated capacity to undergo empirical growth and refinement" (617), through (i) increasing the accuracy of its problem solutions (ii) increasing the number of problems solved and by (iii) extending the problem-solving domain without ad hoc hypotheses.

[6] Another problem is that a theory's rate of progress depends on contingent factors such as the number of scientists being involved in it. If a line of research sees no progress because everyone is occupied with what they perceive to be more fruitful research, that does not imply that that neglected line of research is not promising (see Duerr and Fischer 2025 for details).

*Formal indices* are more closely related to McMullin's notion of U-fertility. They are a "heuristic" that equips a theory with "programmatic research directives" (621) indicating concrete things that scientists can do to increase that theory's performance. Whitt discusses two elements of a theory's heuristic in detail: analogies and experimental strategies.

Whitt specifies an analogy as a "nonliteral similarity comparison between complex systems which maps relations in a known or *base* domain onto a *target* domain of inquiry" (622f, emph. original). An example is Rutherford's analogy between the Solar System and the hydrogen atom. In the Solar System—the base domain—the planets revolve around the Sun. The mass of the Sun is much higher than that of the planets and its distance from the planets much larger than the size of the planets. Mapping these relations onto the hydrogen atom—the target domain—Rutherford could predict that small and light objects revolve around a nucleus that has a much higher mass and is very distant relative to the size of the revolving objects (the electrons).

Whitt identifies four features that constitute an analogy's heuristic power: (i) the analogy's base domain needs to be well understood (*specificity*), (ii) and well established (*validity*), (iii) the mapping of relations between base and target domain should be exact and precise (*clarity*), and (iv) not all the aspects of the similarity should yet be known (*open-endedness*), such that the analogy can be fruitful for generating novel insights. Thus, Rutherford's analogy between the Solar System and the hydrogen atom was a powerful heuristic because Rutherford had an advanced understanding of the dynamics of the Solar System. The underlying gravitational theory was also well established. Moreover, Rutherford could make a clear connection between the respective central objects (Sun, nucleus) and the revolving objects (planets, electrons). Finally, the analogy was open-ended because it indicated a clear research program of what quantities deserved further specification in the atomic model. Importantly, such open-endedness can be exhausted when that research program is concluded. This, according to Whitt's model, is how analogies can cease indicating theory promise.

The second element of a theory's heuristic are its *experimental strategies*. Here Whitt mentions the role of auxiliary hypotheses that are employed to make a theory testable. For example, when Dalton researched atomic weights, he based his experiments on the assumption that chemical compounds are as simple as possible. The promise of such an experimental strategy, according to Whitt, depends on "the variety of evidence that can be cited in its support" (627). Further theory promise arises when scientists can "apply a powerful new experimental device or procedure in empirical problem solving" (627).

Thus, experimentation plays a key role in Whitt's account of theory promise. However, her characterization of experimental strategies remains underdeveloped. What exactly makes an auxiliary assumption fruitful for experimentation? Are there indices that tell us whether an experimental device or procedure is "powerful"? In chapter 3 I will get back to these issues and develop a more detailed account of experimental pursuitworthiness and explain why it matters for theory pursuit.

Summing up, I have discussed three approaches to defining indices of theory promise so far. First, I have discussed Laudan's rate-of-progress criterion, which looks at a theory's recent performance to assess its pursuitworthiness—an approach that faces several difficulties and works only under the assumption that past trends continue. McMullin draws a stricter distinction between the proven fertility of a theory and its future-directed heuristic potential. Finally, by synthesizing insights from both Laudan's and McMullin's approaches Whitt gives the most comprehensive taxonomy of indices for theory promise that I have discussed so far.

Philosophers of science have examined many more indices of theory pursuit, beyond the ones discussed so far. For example, Heather Douglas (2013), takes inspiration from Kuhnian (1977) values of theory choice and argues that "scope, simplicity, and (potential) explanatory power" (800) are relevant values for a theory's pursuitworthiness. She contrasts these values with empirical adequacy and consistency which are more relevant for considerations of theory acceptance. There have also been approaches to developing more fine-grained historical indices. William Anthony Penn and Arica Bauer (2026), for instance, discuss dynamic responsivity—a research program's ability to react to new developments—as a central criterion of pursuitworthiness. Thus, the search for indices of theory promise remains an active field of philosophical inquiry.

Before the discussion on indices of theory promise is concluded, note that all indices that have been discussed so far are construed as global indices. They are thought to be applicable to theory assessment in general, irrespective of scientific discipline and a theory's domain. General applicability is important because it facilitates comparisons between those theories. However, it gives only an incomplete picture of scientific promise. Zooming in on disciplines and specific domains reveals more local indices in addition to such global indices. For example, certain local guiding principles such as the correspondence principle in physics[7] can—among other things—be employed to assess a theory's pursuitworthiness (Fischer, 2023, 2024a).

The discussion so far gives a flavor of the kinds of items that have been identified as indicators for theory promise by philosophers of science. In particular, we now have a better understanding of the positive motivation that can drive scientists to invest their research efforts into theories that are less acceptable than their best-supported competitors.

### 2.3 The Epistemic Status of Promise Reasons

Indices of theory promise give scientists reasons to pursue a theory. If there are sufficiently strong reasons to pursue a theory, those reasons may outweigh the worries resulting from the theory's weak evidential status. If this is the case, pursuit of an evidentially weak theory can be construed as rational. But as agents can have many kinds of goals there are many kinds of rationality. How can we be sure that the kind of rationality that we have recovered here is the right kind of rationality for science? The problem of innovation is a threat to *scientific* rationality. Insofar as science has the goal to improve and increase knowledge this is a kind of epistemic rationality. This kind of rationality should, for example, be distinguished from the rationality that consists in maximizing a research project's economic output, or the personal benefit that a research project can have for the scientist who gains prestige by working on it. Do indices of theory promise promote epistemically rational decisions?

Dunja Šešelja and Christian Straßer (2014) make important progress in addressing this question. They provide a "coherentist account of epistemic justification in the context of pursuit" and, thus, link the debate about indices of theory promise to epistemic justification. They argue that theories are epistemically justified insofar as they exhibit coherence. Coherence amounts to consistency (logical and probabilistic), inferential density, and explanatory power as criteria of coherence (BonJour, 1985). What matters for the acceptability of a theory, according to Šešelja and Straßer, is its *actual coherence*. What matters for a theory's pursuitworthiness is its *potential coherence*. The central analogy here is that of a sketch that is the starting point of a realistic painting of some landscape:

[7] The correspondence principle roughly states that quantum mechanical quantities need to come down to classical quantities in the limiting case of large systems.

> Just like the painter will start with a simple sketch, so does the scientists begin with an abstract model. And just like the sketch is far from being the final form of the painting, ready to compete with other already finished paintings, so is the pursued theory in its beginnings not able to compete with its dominant rival with respect to its actual coherence. Nevertheless, the sketch could already show certain strengths due to which we can say that it seems promising of becoming as realistic a painting of the landscape as other works, if not even more. (3122)

Accordingly, key criteria for a theory's pursuitworthiness are its *potential* consistency, *potential* inferential density and *potential* explanatory power. While consistency in a strict sense is not required for a theory to be pursuitworthy, the more potentially consistent[8] a theory is (both internally and with other theories) the more potentially coherent it is and, hence, the more pursuitworthy it is. A theory's potential inferential density is related to the unified character of its core hypotheses (internal) and its relation to other theories (external). The potential explanatory power is related to the theory's capacity to explain certain benchmark problems in a novel way and to address phenomena that established competitors have difficulties explaining.

For example, Wegener's theory of continental drift was pursuitworthy in the early twentieth century because it fulfilled these criteria (Šešelja & Weber, 2012). Wegener's theory states that all continents had once been part of a single large continent before they separated and drifted apart through the ocean floor. The theory could potentially explain paleontological similarities between the fossils found on different continents and the similarities of structural patterns of the coastlines of matching continents. The basis of this potential explanatory power, moreover, was the single claim of continental drift. Thus, the theory showed a high degree of unification or potential internal inferential density. Finally, since the theory made connections with neighboring fields of physics and seismology it also had high potential external inferential density (see Šešelja and Weber 2012, 152f).

Šešelja and Straßer's account can be seen as building a bridge between indices of pursuitworthiness and a prominent theory of epistemic justification. This approach has clear potential in characterizing promise reasons as epistemic reasons. However, questions arise, as to how far that strategy carries us. Note that Šešelja and Straßer see the need to introduce an additional criterion of "programmatic character" which requires a theory to be embedded "in a theoretical and methodological framework which allows for further research of the system to proceed in spite of the encountered problems, and towards their systematic resolution" (3131). This additional criterion is required to explain why a theory is to be judged pursuitworthy even if in its current form it suffers from significant incoherence (see, ibid.). But the criterion of programmatic character has no equivalent in the coherentist account of epistemic justification. Can it still be counted as properly epistemic? Šešelja and Straßer do not give an answer. Moreover, one might be worried that the requirement of programmatic character stands in a tension with the requirements of (potential) consistency and (potential) inferential density. The more close-knit a set of core hypotheses is, the less breathing space there is for programmatic development.[9]

[8] Šešelja and Straßer employ a graded notion of probabilistic consistency.

What makes an indicator of theory promise an *epistemic* indicator? This depends, of course, on the notion of 'epistemic' that is presupposed here. McMullin and Whitt, for example, both employ a notion that is tied to evidential support. Yet here we are interested in a broader notion of 'epistemic' encompassing things that help us generate and improve knowledge. Will Fleisher (2022) provides a more systematic analysis of such a broader notion of 'epistemic.' He discusses what he calls "inquisitive reasons", that is, "reasons that concern promoting successful inquiry" (24) but that are not necessarily connected with any evidential value.

Consider analogies. As argued by Whitt, they can play an important heuristic role by suggesting new lines of inquiry and empirical tests. But Fleisher points out that these benefits "do not depend on the truth of the associated theory" (2022, 20). Instead, the analogy provides an "extra-evidential favoring" (ibid.) because it may indicate new routes for empirical testing that may turn out in favor of the theory or not. Likewise, a theory's testability promotes successful inquiry because it allows us to determine the theory's empirical adequacy. Testability alone, however, has no evidential value. Similar considerations hold for Šešelja and Straßer's criterion of programmatic character. Programmatic character promotes inquiry by guiding scientists in their research. A theory with programmatic character tells scientists what is to be done to generate and improve knowledge. That programmatic character of itself, however, has no evidential value.

Fleisher argues that to count as inquisitive, promise reasons need to promote inquiry in the 'right' way, they need to be of the 'right kind.' Consider the case of a scientist, Sammy, who likes sandwiches and is more productive when she eats them. Confronted with the decision to join one of two labs *A* or *B* researching two different but equally viable theories, she prefers going to lab *A*, which is located closer to a sandwich shop. According to Fleisher, the "sandwich reason" is not an inquisitive reason because "it does not concern promoting successful inquiry *in the right way*" (26, emph. orig.). The case stipulates that sandwich consumption would increase Sammy's productivity. The availability of sandwiches, thus, would contribute to the achievement of epistemic goals. However, it does so in the wrong way.

[9] Additional issues arise from the *coherentist* nature of Šešelja and Straßer's criteria. Elsewhere (Fischer, 2023) I have argued that there is high potential coherence between the naturalness principle as an important guiding principle in current particle physics and theories of supersymmetry (SUSY). Naturalness and SUSY are consistent with each other, naturalness suggests certain inferences towards SUSY as a possible extension to the Standard Model of particle physics, and SUSY potentially explains why the Standard Model appears to be violating the naturalness principle. Among proponents of naturalness and SUSY this was seen as additional motivation for pursuing SUSY and other solutions to naturalness. Critics, however, see a potentially pathological mutual self-reinforcement of the guiding principle (which has shaky foundations) and the theory (which persistently fails to be confirmed). Coherence alone, it seems, does not help solve this problem. For a more detailed explanation of the Standard Model and physicists' searches for new physics see section 3.1.

Reasons like availability of sandwiches, according to Fleisher are "weak considerations, unlikely to provide a degree of support that makes them worth considering" (26). The cost of including such reasons into decision making, according to Fleisher, unduly complicates decisions. More specifically Fleisher sees a dilemma looming here. If we let our decisions be influenced by all reasons of similar weakness, decision making would be impeded by computational complexity. We simply would not be able to form a decision. If we consider a reason like the availability of sandwiches but no other reasons of similarly low relevance, then we risk giving such a reason inappropriate weight, leading to skewed deliberation.

Even though the availability of sandwiches increases epistemic output it does not represent an epistemic reason to prioritize pursuit of theory *A* over pursuit of theory *B*. What is instructive for our purposes is that Fleisher dismisses this reason because he thinks it is too weak. The availability of sandwiches has only a comparatively small effect on Sammy's performance and that preference for sandwiches is idiosyncratic to her. Filling a research campus with sandwich shops of the kind that Sammy prefers, presumably, is not a good strategy for promoting epistemic goals, because the difference that the availability of such sandwiches makes is comparatively small and potentially specific to Sammy.

However, if a reason like the sandwich reason becomes sufficiently weighty, it will, by Fleisher's line of reasoning, become the 'right kind' of reason to be considered for epistemic purposes. For example, logistic considerations like the availability of food on campus may under relevant circumstances be important for substantially increasing researcher productivity. By promoting goals of inquiry such external considerations, thus, gain the status of inquisitive reasons and will count as epistemic.[10]

Going back to Sammy's decision to work on theory *A* or *B*, however, this result needs to be qualified. Even in the altered case where lab *A* is equipped with food logistics and lab *B* is not, it is—in a certain sense—odd to state that for this reason theory *A* is more pursuitworthy than theory *B*. The reason is that availability of food at lab *A* is not an intrinsic property of theory A. Closing food stores at lab *A* would not make theory *A* less attractive as a subject of scientific interest. What needs to be distinguished here is the pursuitworthiness of the theories *A* and *B* themselves and the pursuitworthiness of the wider research contexts of *A* and *B*, which include the theories but also the practical issues that enable and promote research. Importantly, however, that distinction is not always drawn easily, especially when we are concerned with the pursuitworthiness of experimentation.

[10] Šešelja et al. (2012) agree with this result. Such external considerations matter for epistemic pursuitworthiness in so far as they "are in function of epistemic goals" (67).

Summing up, promise reasons are epistemic reasons and thus help us restore the epistemic rationalist of scientific pursuits. But to achieve this, promise reasons need to promote inquiry in the right way. Factors that bring about only marginal increases of epistemic gains do not qualify because they would either bias decision making or make decision making overly complex. Moreover, we should distinguish between promise reasons that are tied closely to a theory, such as a theory's analogies, and promise reasons that are more contextual, such as the availability of resources that enable and promote work on a theory. In what follows, these more contextual promise reasons will be of special relevance when I move on to discuss the pursuitworthiness of experimentation.

### 2.4 Pursuitworthiness and Scientific Decision Making

Now we have a better understanding of why it can be rational to pursue a theory that is less acceptable than its strongest competitors. But scientists' decisions often need to be more specific. They need to decide which one of the many available theories should be pursued. The foregoing discussion does not carry us very far when it comes to such decisions. Consider Laudan's rate-of-progress criterion. The criterion assumes that one can count and weigh the relevance of problems. But how are we supposed to count problems? More critically even, if there is disagreement about the pursuitworthiness of a research program, we also have to expect that there is disagreement about how relevant the problems are that the research programs purport to solve. If there is no independent measure for assessing number and relevance of problems, Laudan's criterion will not help us solve any disagreement about what path of research is to be followed. Unfortunately, the situation does not get much (or any) better if we move to other indices, such as the availability of analogies and experimental strategies. These indices are too abstract to guide decision making.

So, does philosophical thinking about pursuitworthiness have anything to offer for scientists and policy makers seeking advice on what lines of research to pursue? Responding to this question Jamie Shaw (2022) extends Feyerabend's worries, arguing that even minimal criteria for selecting pursuitworthy research projects are hard to establish. More specifically, he analyses the viability of what he calls the "Equivalence Principle: We can treat all research proposals as equally viable forerunners of success" (106). The principle states an equivalence of the scientific promise associated with all available research proposals at a time. The consequence of the equivalence principle is that there is no reasoned choice to be made between research projects. Choices "will be a matter of taste" (106).

Shaw endorses the equivalence principle but argues that it has limited scope. He argues that it holds for what he calls "luxury science" while it does not apply to what he calls "urgent science". Urgent science, according to Shaw, is research that has a timeline imposed by practical or moral needs. For example, vaccines can be urgent when the vaccine is needed to immunize a population against a fast spreading and aggressive virus, such as Covid-19. "Luxury science," according to Shaw, is research that has no such timeline because there are no expectations for its practical application. Shaw's examples for luxury science are cases of foundational research such as experimental tests of Hawking radiation.

Shaw also provides empirical evidence in support of his claims looking at predictive validity studies (PVS). PVS assess correlations between the review scores of article manuscripts or research proposals and their down-the-road success, measured, for example by citation scores or policy uptake. Shaw argues that the predictability of research success is given to some degree for the short term (less than 5 years) but not for the long term. He takes this to support his claim that the equivalence principle holds in "luxury science" but not in "urgent science".

The classification of certain kinds of foundational research as "luxurious" may be understood as coming with an implicit evaluation of that science as expensive and useless, like jewelry. But this is evidently not what's meant here, when "luxury science" is introduced as the contrast class of "urgent science": there are many instances of science that are not strictly urged by moral or practical needs yet are clearly useful for scientific progress and comparatively cheap, such as most theoretical work in a discipline like physics.

The distinction rather concerns the timeline on which research is conducted. On the one hand, there are instances of research, especially applied research, that directly respond to epistemic and technological needs. And these needs can be coupled with strict timelines dictated by external events such as the outbreak of a pandemic. On the other hand, there are instances of research that do not (or are not expected to) respond directly to such needs, but advance science on more foundational levels. The benefits of society from such projects are typically much less foreseeable and may be realized only further down the line.

But does that imply that instances of foundational research have no timeline? A key assumption underlying Shaw's distinction is that there are no purely epistemic reasons for why scientists should prefer to have a research result rather earlier than later. If a project's output is not applied to address a societal need with a pressing timeline, then, according to Shaw, there simply is no timeline. He concedes that scientists in foundational research may have strong interests in timely scientific discoveries but argues that this is for purely non-epistemic reasons such as "promotion and prestige" (107).

However, it is questionable whether (i) such a clear line between epistemic and non-epistemic reasons can be drawn here, and whether (ii) timelines really are not relevant for purely epistemic reasons. Suppose a research project has a long timeline before any significant results can be expected. This will make that research project less attractive for researchers because they cannot hope to receive "promotion and prestige." If attractiveness goes below a certain threshold, there will simply be no researchers advancing that project such that the project will eventually come to an end. But this implies that that project's potential results will not simply be delayed but may potentially *never* be realized, meaning that there will be net epistemic loss. A related concern comes up in particle physics, arguably, a prime example of what Shaw calls luxury science. For example, there are worries that discontinuing the series of large experiments as part of the European particle physics strategy will have lasting effects on this community's ability to achieve further epistemic goals. If there is no follow-up machine to the Large Hadron Collider (LHC) what experiments are young particle physicists to be trained on? Thus, discontinuing or even just delaying a line of research for some time can have significant costs for a research community that will eventually impede that community reaching also its purely epistemic goals. Thus, timelines—especially science-internal timelines—will also play a crucial role in foundational research.

Shaw argues that the equivalence principle holds for projects without a clear or short-term timeline, meaning that it is hard to establish even minimal criteria of pursuitworthiness for such projects. However, here we have seen that also primary instances of foundational research that do not respond directly to science-external needs have clear (internal) timelines. Does this mean that, *pace* Shaw, the equivalence principle does not hold, even in its restricted form?

It certainly is harder to predict the success of foundational research—but this is not necessarily related to timelines but rather to whether research is performed in direct response to a well-articulated need or not. If there is such a well-articulated need, then projects can be clearly defined such that a solution can be developed and evaluated in a stepwise and well-organized fashion. The outputs of foundational research, however, are often less predictable than those of applied research. For example, the likelihood that a research project must be altered during pursuit is higher, because of unforeseen interim results.

In summary, Shaw extends Feyerabendian worries that no minimal criteria can be given for governing scientific pursuits. More precisely, he employs a distinction between what he calls luxury science and urgent science and argues that minimal criteria for pursuitworthiness may be given only for urgent science, where clear timelines are dictated by science-external needs. Here I have challenged Shaw's distinction between luxury and urgent science. Given that also foundational research is governed by strict timelines one should expect that there also arise clear demands on criteria for deciding about scientific pursuits.

### 2.5 Criticizing Pursuits

The lack of foreseeability, especially in foundational research, will make it difficult to provide strict criteria for *ex-ante* pursuitworthiness. However, it does not mean that no overarching guidance can be given by philosophical or methodological reasoning. Before we conclude this chapter, we will look at two recent frameworks for such methodological guidance. Guidance through criticism and guidance through economic reasoning.

We need concrete guidance for critically assessing specific scientific pursuits. Marina DiMarco and Kareem Khalifa (2022) put such criticism at the center of their model. What matters for the evaluation of pursuits, according to DiMarco and Khalifa, are what they call apocritic norms. Apocritic comes from the Greek *apokrino*, which means 'to give an answer.' Apocritic norms are norms that govern practices of giving answers to questions.

Apocritic norms come in two forms: obligations and prohibitions. An apocritic obligation has the form: "If a question Q about object of inquiry x has feature F, then some scientists with capability C should pursue question Q about x" (87). Suppose humanity is threatened by a global pandemic. Suppose also that there are scientists with the capability to research vaccines that could help prevent the pandemic. Then the apocritic norm would act as a basis for criticizing those scientists if they did not pursue research on that vaccine. An apocritic prohibition takes the form: "If a question Q about an object of inquiry x has bug B, then no scientist should pursue question Q about x" (87). For example, if some experiment leads to the unnecessary suffering of lab animals, then that experiment can be criticized based on the apocritic prohibition that such unnecessary suffering should be avoided.

DiMarco and Khalifa's model gives a framework for criticizing concrete pursuits. The model also highlights the many potential aspects of such criticism. In particular, it moves extant discussions forward by relativizing pursuitworthiness to the capacities of concrete scientists, such as the expertise needed to address a research question. Moreover, the model gives epistemic reasons a role in the evaluation of scientific pursuits, but it also acknowledges that epistemic reasons will only be part of any such evaluation, as non-epistemic features and bugs are to be taken into consideration as well.

There is, however, a question of how specific the obligations and prohibitions should be, and how strongly they can and should constrain scientific decision making. If obligations and prohibitions concern only extreme cases such as fighting a pandemic and the prevention of unnecessary animal suffering, the norms will leave concrete issues of pursuitworthiness heavily underdetermined: there are many things a scientist can do to help fight a pandemic. Will there be a norm telling them what concrete theory to pursue? Providing overly specific obligations and prohibitions, on the other hand, could bear the risk of placing overly strict and potentially conflicting constraints on scientific decision making. What should scientists do when the only available method to develop vaccines requires harming lab animals?

The latter concern is to some degree alleviated because DiMarco and Khalifa characterize apocritic norms as *pro tanto* obligations and prohibitions. This means that, all things considered, these norms can be overridden. For example, the obligation to address a research question may be overridden if another research question comes up that is even more urgent. Likewise, prohibitions may be overridden if certain features obligate the pursuit of a project. However, this raises additional questions of what the higher-order norms are that should govern such overriding. Most research projects come with useful features that demand pursuit *and* bugs that would speak against pursuit. The important question for pursuitworthiness is whether the useful features outweigh the bugs.

### 2.6 Pursuitworthiness and the Economy of Research

Decision-theoretic or economic approaches implement such weighing of advantages and disadvantages from the get-go. The starting point of these approaches is Charles Sanders Peirce's idea of an *economy of research* (Rescher, 1976). Peirce (1960) observes that "[p]roposals for hypotheses inundate us in an overwhelming flood, while the process of verification to which each one must be subjected before it can count as at all an item, even of likely knowledge, is so very costly in time, energy, and money […]" (5.602). The premise of economic approaches is that there are many more possible lines of research than can actually be pursued. The task is to select from these the lines of research that promise the largest return on investment. Thus, the basic idea is that of a cost-benefit analysis.[11]

Since our focus here is epistemic pursuitworthiness, the benefits are epistemic benefits, that is, whatever increases or improves knowledge. In discussions of theoretical pursuitworthiness this amounts to new theories and models, novel hypotheses, and theoretical guiding principles, elaborations of theories and models that help us see the theories' empirical predictions or provide a better understanding of empirical phenomena. In chapter 3 we will see that the concept of epistemic benefit can easily be extended to also include gains that are not of a primarily theoretical nature (e.g., new data) and may have various kinds of indirect epistemic value (e.g., training new lab personnel). The costs reflect the efforts in developing a theory. In theory pursuit this primarily amounts to cognitive efforts. But evidently, these are only a small part of what is required to establish a theory. A large part of the costs, especially non-cognitive ones, are related to empirical tests of theories. This is another important reason why it will be instructive to look more closely also into the pursuitworthiness of experiments as will be done in chapter 3.

The pursuitworthiness *PW* of a research project *R* is a function of that project's expected epistemic gain (*EEG*) and the associated costs (*C*):

$$PW(R) = EEG(R) - C(R).$$

The expected epistemic gain, in turn, depends on the epistemic gain associated with the project's potential outcomes and the likelihoods of achieving those outcomes, given that the project is pursued. For example, if the project addresses a hypothesis *H*, then the expected epistemic gain depends on the epistemic gains associated with accepting or rejecting (or withholding judgment on) hypothesis *H* weighed by the probabilities of these outcomes (Nyrup, 2015, p. 755).

The formula gives the impression that precise pursuitworthiness scores can be calculated. But it should be emphasized that any attempt to perform such calculations is confronted with principled limitations. In particular, calculations would require that the expected epistemic gain associated with a research project can be expressed in the same 'unit' (e.g., dollars) as the costs associated with the research effort. Moreover, calculation of the expected epistemic gain is possible only in so far as all the research project's possible outcomes are known and precise likelihoods can be assigned to them. Whether this is possible is questionable, and especially so in the context of foundational research.

[11] For a detailed account of the historical origins of Peirce's economy of research in the context of his work as a government-funded geoscientist see a recent article by Alisa Bokulich and Matthew Brewer (2026). Bokulich and Brewer argue that the target of Peirce's account of pursuitworthiness were not hypotheses or theories but data.

But even if such concrete calculations cannot be performed, the approach still facilitates certain comparisons. Suppose two projects *R1* and *R2* have the same expected epistemic gain but *R1* is associated with higher costs. Then the approach recommends pursuing project *R2*. Likewise, if projects *R1* and *R2* have similar costs but *R2* has a higher expected epistemic gain the approach gives a clear recommendation for *R2*. The difficult decisions, of course, are the ones in which some extra expected epistemic gain is to be traded off against additional costs.[12]

So, if concrete calculations are hard to make, what is the purpose of the economic approach? In joint work with Patrick Duerr (Duerr and Fischer 2025) I have described the economic model as a meta-methodological framework for assessing considerations of pursuitworthiness. Unless one provides a more substantial account of costs and benefits,[13] the framework is not particularly informative as a guideline to concrete decision making. But it gives guidance for assessing methodologies of pursuitworthiness on a meta-methodological level. For example, the approach can explain the important role of analogical reasoning for considerations of pursuitworthiness. Rune Nyrup (2020) argues that "analogies facilitate the transfer of an already well-understood modelling strategy to a new domain" (897). Thereby analogies can reduce the cost of developing a theory of that new domain. What matters in this context, according to Nyrup, is understanding-with. Understanding-with is the kind of understanding that one needs to generate explanations on the basis of a model or theory and it is the precondition for understanding-why, which is concerned with phenomena. Analogical reasoning can transfer understanding-with from the base domain (which is typically well-understood) to the target domain. Thereby it lowers the costs of pursuit, because understanding-with in the target domain does not have to be generated from scratch. Here the economic framework does not predict any concrete savings from such transfer. Yet, on the meta-methodological level incorporating the notion of 'cost' helps explain why analogical reasoning can play such an important role in the context of pursuit.

[12] There are additional strong assumptions here (1) that such comparisons of epistemic gain can in principle be made, and (2) that there is sufficient agreement about the assignment of such epistemic gains among proponents of potentially competing research projects.
[13] Duerr and Fischer (2025) provide such an account in terms of theoretical virtues, more on this below.

Moreover, earlier we saw that DiMarco and Khalifa relativize obligations and prohibitions to scientists' capabilities. From the perspective of economic models this is important. How costly a research project is, will depend on the capabilities and infrastructure that scientists bring to the project. However, it is an advantage of the economic model that cost is the more general factor than present capabilities. Consider a case where no scientist yet possesses the capabilities to tackle a research question. Here, pursuitworthiness hinges on the cost of acquiring those capabilities.

The economic framework is also a useful heuristic device for structuring discussions of pursuitworthiness. More specifically, the analogy to economic decision making raises a series of questions that guide the meta-methodological assessment of pursuitworthiness:

- What are the epistemic benefits that are being promised if an idea is pursued?
- What are the costs incurred by pursuing an idea?
- Who is evaluating the pursuit?
- How are costs and benefits to be weighed?

Insofar as the economic model is understood as a meta-methodological framework, there are various ways of answering these questions in a coherent way. Duerr and I (2025) have looked at the case of theory pursuit from the perspective of a suitably idealized scientist that seeks to promote epistemic goals. From this perspective benefits and costs are assessed based on theoretical virtues, such as empirical adequacy, consistency, scope, and simplicity (see also Douglas, 2013; Keas, 2018; Kuhn, 1977) that are being weighed in a deliberative process. Alternatively, one may take a less idealized perspective that will take into consideration non-epistemic reasons for pursuit and contextual factors such as the cognitive, organizational and material resources that are already in place in a lab or group of scientists.

Summing up this and the foregoing section, we have seen two positive approaches to decision making about pursuit. One approach is to characterize pursuitworthiness criteria as context-sensitive and person-relative pro tanto prohibitions and obligations, as suggested by DiMarco and Khalifa. However, such prohibitions and obligations always need to be weighed against each other when concrete pursuits are evaluated. Economic approaches incorporate such trade-off relations from the get-go by conceptualizing them in terms of (epistemic) benefits and costs. This provides a clearer view of the reasons why certain indices such as the presence of analogies promote pursuitworthiness. A central challenge for this last approach, however, is to spell out what these benefits and costs amount to in concrete instances and how they are to be weighed.

## Chapter 3: Promising Experiments

For Laudan the primary objects of appraisal are research traditions. Research traditions include methodological commitments and, arguably, experimental practices and strategies. However, as the survey in chapter 2 illustrates, subsequent philosophical debate has been predominantly concerned with the pursuitworthiness of theories and the conceptual part of research traditions, while experimentation has been largely neglected. This chapter will provide a new perspective by arguing that the pursuitworthiness of experimental practices and strategies plays a central role as well (section 3.1). After a brief look into the epistemology of experimentation (section 3.2) and a few clarificatory remarks on the context of pursuit in experimentation (section 3.3) the chapter will present an account of experimental pursuitworthiness (section 3.4). More specifically, I will employ the economic framework presented in chapter 2 to provide a structured discussion of a series of factors that do and should impact assessments of pursuitworthiness, such as an experiment's potential epistemic gains, the expected costs, the ex-ante probability of the existence of the experiment's intended outcome (e.g., a new phenomenon), and the experimental strategy's capacity to achieve that outcome given that it exists. A central role in such weighing, it will be argued, is played by criteria that can be characterized as experimental virtues. Finally, section 3.5 will round off the discussion by giving special attention to the various kinds of uncertainty that are associated with assessments of experimental pursuitworthiness, and what strategies scientists can employ to manage such uncertainties.

### 3.1 The Importance of Experimental Pursuitworthiness

The costs of empirical pursuits can be enormous and often exceed the costs of theory development by far. This, by itself, however, is not a sufficient reason for looking into the pursuitworthiness of experiments, one might think. Maybe the pursuitworthiness of experiments just is—or comes down to—the pursuitworthiness of the theories that those experiments address. Let us call this the reductionist assumption. If the reductionist assumption holds, then the topic of scientific promise would be covered exhaustively by a discussion of theory promise. Some of the literature surveyed in the foregoing chapter can be seen as implicitly invoking this assumption because of its strong focus on theory promise and neglect of experimental promise.

But the reductionist view is untenable for at least two reasons. First, theoretical pursuitworthiness depends on experimental pursuitworthiness. Key criteria governing the context of theory pursuit refer to the theory's empirical problem-solving capacity. To understand that capacity we need an analysis of the empirical methods that are being employed. And since we are concerned with the context of pursuit, we need to look at what makes these methods pursuitworthy. Second, it is a philosophically well-rehearsed point by now that experimental practice is to a certain degree independent from theoretical work and often gives rise to hypotheses and concepts in the first place. Consequently, one should not assume that the criteria that hold for successful theory pursuit simply carry over into the case of experiment. I will spell out both points in more detail.

First, note that key indicators of scientific promise refer to empirical methods. For example, Laudan's (1977) rate-of-progress criterion refers to problem-solving of both theoretical and empirical problems. To solve empirical problems, however, a theory needs to be put in touch with the empirical. A theory's ability to address empirical problems does not only depend on that theory's predictions but also on scientists' abilities to test those predictions. Likewise, Whitt (1992) argues that pursuitworthy theories are accompanied by "experimental strategies" that need to be supported empirically or by a "powerful new experimental device or procedure in empirical problem solving" (627). But what does such power amount to? Can we develop further indices for such power? And what does the relation between theory and experiment have to look like such that empirical fruitfulness is established?

The dependence of theoretical pursuitworthiness on experimentation is particularly evident in contemporary foundational physics, where theoretical proposals abound but experimental methods face severe limitations in providing tests of these proposals. Correspondingly, there has recently been a surge in case-study-based philosophical discussions of pursuitworthiness covering fields such as String Theory (Cabrera, 2021; Camilleri & Ritson, 2015; Ruiz de Olano, 2023), Beyond the Standard Model physics (Chall, 2020; Fischer, 2024b; King, 2025), and cosmology (De Baerdemaeker & Boyd, 2020; Wolf & Duerr, 2024).

To get a flavor of the issues raised here, consider the example of particle physics. The Standard Model (SM) of particle physics is currently our best foundational theory of the physics of matter. It classifies all known fundamental particles and provides a unified description of the electromagnetic, weak, and strong interactions—three out of four of the known fundamental forces (with gravity being the fourth that is not accounted for). Its predictions are also of unprecedented precision. At the same time, the Standard Model is known to have a variety of shortcomings. This is why physicists are looking for phenomena that are not captured by the Standard Model, phenomena that would indicate a theory "beyond" the Standard Model, hence "Beyond the Standard Model Physics", or for short, BSM physics. More specifically, the Standard Model is assumed to be predictively adequate for particle collisions up to a certain energy level, while it is thought to fail or break down if it is employed to describe processes taking place at higher energy collisions. A central question for particle physics (but certainly not the only one) is: where on the energy scale is that point of breakdown? Now, ‘breakdown’ sounds dramatic. But in fact, finding a breakdown is what physicists very much hope for, because it would give indications for novel phenomena (Ritson, 2020)—the kinds of discoveries that would earn particle physicists a Nobel Prize.

The Standard Model is expected to break down when it comes to predictions at arbitrarily small distance scales or arbitrarily high collision energies (which in particle physics amounts to the same). The reason is that the Standard Model does not include a description of gravity, which becomes relevant at those small distances or high energy scales. However, if we wanted to probe that energy scale experimentally with current technology, we would need a particle collider of the size of $10^{10}$m, about a tenth of the distance between Earth and Sun (Zimmermann, 2018). This is, without doubt, a line of research that does not sound very promising from the viewpoint of the experimentalist.

But the lack of gravity is not the only shortcoming of the Standard Model. Consider the Higgs boson. Even though the boson was discovered only in 2012, particle physicists have been discussing since the late 1970s about how to explain the mass of that boson (e.g., Susskind, 1979). It seems that to explain the mass that was confirmed experimentally in 2012, the parameters of the Standard Model need to be very finely tuned—a fact that does not represent a straightforward inconsistency of the model, but a fact that many physicists thought cries out for a deeper explanation. Such fine tunings, many physicists believed, would not be required if the Standard Model’s point of breakdown is located at an energy regime fairly close to where the Higgs boson was discovered.

This argument from fine tuning—also called the naturalness argument (Williams, 2015)—was a driving force in experimental particle physics over recent decades. It was taken to be a strong argument to the effect that novel discoveries are in reach of current or upcoming colliders. Only the persistent absence of any new discoveries since the discovery of the Higgs boson has cast doubt on this argument (Fischer, 2023). And now with particle physics having entered what has been called the post-naturalness era (Giudice, 2018) physicists are much more pessimistic regarding such discoveries (Harlander et al., 2023).

Looking at this recent episode of research in foundational physics is instructive for our purposes because it shows that the pursuitworthiness of theories is strongly coupled to our capacities to perform empirical tests on them. The key role of fine-tuning arguments in the recent dynamic of the field was to nurture hopes that empirical advancement is on the horizon (Fischer, 2024b). Even since naturalness and related thoughts have lost traction considerations of testability play an important role in particle physics model building. In particular, it has been argued that the demand for testable models has led the field to move away from "the pursuit of beautiful, simple, and general theories" towards the pursuit of theories that are "ad hoc, narrow in scope, and complex" (King 2025, 1). But if the pursuitworthiness of these models depends on their testability, it will be necessary to say more on the conditions that make such tests pursuitworthy. That we cannot build a collider of the size of the Solar System is clear. But presumably there is much more to be said about what kinds of instruments are feasible and worth the effort.

So far, I have argued that the pursuitworthiness of theories depends on the pursuitworthiness of certain experimental practices. But there are additional reasons for taking experimental pursuitworthiness seriously. In particular, experimental practice is to a certain degree independent from theoretical work and often gives rise to hypotheses and concepts in the first place. That partial independence from theory takes many forms. One straightforward aspect of it is that often there is no one-to-one mapping between theoretical research questions and the experiments that are employed to address them (see Fischer (2026b) for an extended argument along these lines).

First, for any theoretical research question there is often more than one way to address it. That is, once we have decided that a theoretical question should be addressed, that still leaves the concrete experimental procedure to be followed and the experimental facilities to be set up widely underdetermined. There arise additional questions as to what procedures and facilities are the most pursuitworthy ones to address the theoretical question. For example, the decision to probe particle collisions at higher energy levels leaves open to a certain degree the concrete collider design that is most suitable for such experiments.

Second, often experiments are not performed for the purpose of addressing a single research question. Under these circumstances an experiment's pursuitworthiness exceeds the pursuitworthiness of any one of the individual theoretical research questions that is being addressed. Sometimes, of course, an experimental facility is put in place to address just one question, and such pursuit may well be justified. More generally, however, it will be a matter of good lab management to choose instruments that can be reused for a variety of purposes. And under such circumstances the value of an instrument does not come down to any individual research question that are being addressed but rather to combinations of research questions. This applies in particular to instances of 'Big Science,' where experiments are developed to support whole lines of research (Shaw, 2025; Vijay & Arrabito, 2026).

Another important point illustrating that experimental pursuitworthiness is partially independent from theory is that experiments and experimental methods can give rise to new lines of research once they are established. Examples can be found in what Siska De Baerdemaeker (2021) has described as method-driven experimentation. Common experimentation is target driven. Here the choice of the experimental method is justified by assumptions about the experimental target. However, sometimes knowledge of the experimental target is so limited that such justification is not viable. For example, in searches for dark matter the target-driven approach is limited because only very little is known about the properties of dark matter. In such instances scientists sometimes perform method-driven experimentation. Method-driven experimentation also requires certain basic assumptions about the experimental target but primarily "looks at the method for guidance as to what further assumptions about a target system might need to be made in order for a method to be effective to discover new features of the target system" (142). The main justification for an experimental strategy does not derive from the theory of the target but from the knowledge about the available methods. For example, dark matter searches at the Large Hadron Collider are not primarily motivated by specific assumed characteristics of dark matter (which we know very little about) but by the assumption that *if* dark matter had certain characteristics, *then* it would be detectable by collision experiments.

In summary, the pursuitworthiness of experimentation merits further discussion. Not only does theoretical pursuitworthiness evidently depend in important ways on the pursuitworthiness of experiments. We also should expect that experimental pursuitworthiness raises a series of issues that remain invisible if we restrict philosophical attention to theory pursuit.[14]

### 3.2 The Philosophy of Experimentation

Let us have a look into the philosophy of experimentation before we dive into questions of experimental pursuitworthiness.[15] Empirical practices are a central element of science. Scientists often consider empirical evidence the ultimate criterion to evaluate a theory. The relevance of the empirical is also part and parcel of central views in the history of philosophy of science. Consider the positivist principle of verification that demands that every meaningful statement be either empirically verifiable or to be an analytic truth. Or consider the falsificationist requirement that every theory be falsifiable. Thus, theory is worth little if it cannot be put in touch with the empirical.

[14] Let me emphasize that I am *not* claiming that experimental pursuitworthiness is entirely independent from theory pursuit. In fact, most of the examples discussed here and below are still in some way or another motivated by theory. Consequently, an experiment's relation to theory will be essential in understanding that experiment's pursuitworthiness. For an example in which experimental pursuitworthiness exhibits a particularly far-reaching independence from theory see Bokulich and Brewer's (2026) discussion of the pursuitworthy data collection in the geosciences.

[15] The survey here will be brief. For an excellent introduction to the topic see Boyd (2021).

Among the many empirical strategies that scientists pursue, experimentation is typically considered a particularly powerful method.[16] Adrian Currie and Arnon Levy (2019), for example, argue that experiments have particular epistemic power if understood as "*controlled* investigations of *specimens*" (1066). A controlled investigation is one that usually isolates the object of study to a high degree from external influences, e.g., by performing the study in a laboratory environment. Typically, it also implies that the object of study can be manipulated in a fine-grained way and with regard to its properties of interest. Moreover, such experimental procedures can ideally be repeated and are concerned with an aptly chosen specimen that is representative of the class of objects of interest.

Over long stretches in the history of the philosophy of science experimentation was considered to be just that: a particularly powerful testing ground for theory. Since the 1980s there has been an increased awareness that experimentation has more than this auxiliary role in scientific methodology. Experimentation "has a life of its own" (Hacking, 1983): it often develops independently from theory and generates new ideas and hypotheses. With this came also a methodological shift in the philosophy of science: it is no longer sufficient to look at the results of experiments, but detailed case-study-based research on experimental practices is required.

With a more detailed look at experimental practices came also a new awareness (1) of the epistemic challenges faced by experimentation and (2) of the variety of experimental practices. In the remainder of this section, I will address these two points in turn and will discuss their implications for the pursuitworthiness of experimentation.

The replicability of an experiment is central for its acceptability. If a scientist claims to have found a general truth about their research object, their experimental results should be reproducible beyond the doors of their lab. The key role of replicability for science has only recently been highlighted in the replication crisis in psychology. While this crisis can be understood as referring to many issues such as the lack of sufficient replication studies, the evidence of publication bias and a prevalence of questionable research practices, it is especially the widespread failure of replication studies that raises epistemological concerns (Fidler & Wilcox, 2026).

Well before the replication crisis in psychology became a topic of epistemological concern, sociologist of science Harry Collins (1985) performed detailed studies on replicability in physics. Looking at the construction of TEA lasers (transversely excited atmospheric lasers), he observed that it is very hard or even impossible to guarantee the replication of complex experimental procedures. What is required for successful replication, according to Collins, is the transfer of skill-like capacities that travel only through apprenticeship, which requires direct contact between scientists.

[16] See Okasha (2011) for a systematic argument for the epistemic superiority of experiments over other empirical methods. See Matthiessen and Boyd (2024) and Boge (2025) for alternative takes on this issue.

The haphazard character of replication becomes problematic when there is scientific disagreement about the existence of the target phenomenon. In the 1970s physicist Joseph Weber claimed to have discovered gravitational waves, ripples in spacetime predicted by Einstein's theory of general relativity. When other physicists failed to replicate Weber's result, Weber argued that those attempts at replication were unsuccessful because they were not performed correctly. In hindsight we know that Weber's measurements did not track gravitational waves. Direct detection was only possible with the help of large interferometers in 2015.

The episode of scientific disagreement in the 1970s prompted Collins to put forward the concept of the experimenter's regress, which describes a vicious circularity of justification. To validate the purported phenomenon, one needs to refer to the experimental method that purports to establish that phenomenon. Since gravitational waves were a novel phenomenon, the only available experiment was that by Weber. To validate the experimental method, however, one needs to refer to the measured phenomenon (gravitational wave). Any attempted replication that did not track the supposed signal of gravitational waves was simply rejected by Weber as invalid. In such a situation, disagreement about phenomenon and validity of experiment are inextricable. For Collins, a proponent of the 'Strong Programme' in the sociology of scientific knowledge, this indicates that scientists make use of science-external evaluation criteria to resolve such disagreement.

This line of reasoning was forcefully countered by Allan Franklin's (1989) epistemology of experiment. With a close eye to a large number of examples, mainly but not exclusively from particle physics, Franklin has developed a list of strategies that experimenters can employ to counter issues of the kind described by Collins. For example, by calibrating an instrument on a phenomenon that is well understood scientists can increase the trust in this instrument to also be able to detect yet unconfirmed phenomena.

What matters for our purposes is that these discussions have focused almost entirely on what could be called the "context of acceptance" of experimental results. Franklin asks what justifies the acceptance of a scientific result, and he is concerned with "how we come to believe rationally in an experimental result" (1989, 437). The pursuitworthiness of experimentation, however, has barely been an issue in these discussions.[17] Only more recently Franklin and Ronald Laymon have addressed the context of pursuit of experiments. In their book *Case Studies in Experimental Physics: Why Scientists Pursue Investigation* (2022) they survey a wide range of experiments that physicists have deemed pursuitworthy. Laymon and Franklin (2024) also summarize a few lessons to be drawn from such case studies. For example, they observe that an experiment "may be deemed pursuit worthy even in the face of successive failures if it is judged to be the only possible type of approach for dealing with a significant theoretical question and there is some evidence to support optimism as to experimental improvements" (23). Laymon and Franklin's observations are valuable and well-connected to specific examples. However, by saying what "may" lead to pursuitworthiness they make a very weak claim, and they do not provide an overarching framework for thinking about the pursuitworthiness of experiments.

Another central lesson of philosophical and historical scrutiny directed at experimentation is that experimentation serves a wide range of purposes apart from testing theories. Experiments can stimulate new theories, as in the case of optical experiments with Iceland Spar (Hacking 1983), and they are performed to characterize novel phenomena, such as in Röntgen's early experiments on x-rays. Experiments are also performed to establish new measurement techniques and instruments such as in the development of thermometry (Chang, 2004), and they can provide a proof-of-concept, such as in the engineering of cyanobacteria for biofuel production (Kendig, 2016). Moreover, experiments are not only performed to generate new insights and methods, but also to replicate or challenge extant experimental outcomes (see above), or to educate and train students.

Keeping track of these various goals of experimentation is crucial when looking at experimental pursuitworthiness.[18] For example, experiments that are performed for purposes of education are particularly pursuitworthy when they advance the understanding of the subject matter of those who observe them or are instructed to perform them. This is, for example, advanced if the experiment has perceptible outcomes and an element of interactivity. Moreover, experiments for educative purposes should require only simple material and should be safe to perform in a classroom or student lab. Perceptible outcomes, the use of only simple material and safety are virtues that will also count in favor of an experimental setup that is being employed in non-pedagogical circumstances, e.g., in theory testing that is performed to advance cutting-edge research. The point is that under such circumstances one is likely to trade such virtues for virtues that matter more directly for theory testing such as a better signal-to-noise ratio or additional precision.

[17] This is unfortunate: issues of pursuitworthiness may play an important role in understanding how disagreements, e.g., regarding Weber's claims are resolved. Presumably, at some point physicists did no longer take it to be pursuitworthy to try to replicate Weber's results.
[18] For a more detail argument for this claim see Vanessa Seifert's (2026) discussion of the pursuitworthiness of experimentation in chemistry.

A distinction that is particularly interesting for our purposes is that between hypothesis-testing and exploratory experimentation. Experimentation often aims at testing hypotheses, such as the Michelson-Morley experiment that tested the hypothesis of an ether drift. However, there are also instances, in which no specific hypothesis is at stake, and that have a more open and exploratory character. As an example, consider Ampère's early work on electromagnetism, as discussed by Friedrich Steinle (1997, 2016). In 1820, Oersted had discovered that the connecting wire of a voltaic battery had an impact on a magnetic needle located nearby. Oersted's observations prompted Ampère to analyze the impact in more detail. By systematically varying the relative position of a magnetic needle, Ampère could establish that the needle swings into a position perpendicular to the direction of the wire. Importantly, Ampère did not perform this experiment to confirm or reject an extant theory or hypothesis, because there was no such theory or hypothesis when he started his work. Instead, the systematic variation was performed with the goal of finding a general rule in the first place. Thus, Steinle describes exploratory experimentation as a procedure that involves varying a large number of different parameters with the primary goal of establishing new concepts and regularities. Steinle's proposal focuses on case studies in the history of electromagnetism. Others have identified similar episodes of exploratory experimentation in molecular biology (Burian, 1997; L. R. Franklin, 2005; O'Malley, 2007), nanotoxicology (Elliott, 2007), and particle physics (Beauchemin & Staley, 2024; Karaca, 2013, 2017; Mättig, 2022). These studies have also led to a more comprehensive understanding of the various forms that exploratory experimentation takes, and there are now detailed accounts available about how to categorize them, such as Kevin Elliott's (2007) taxonomy that distinguishes exploratory experimentation along three dimensions: (1) the aims of experimental activity, (2) the role of theory, and (3) the methods or strategies for varying parameters.

So, suppose that exploratory experimentation has a set of goals that distinguishes it from theory testing experimentation. Does that have implications for the kinds of characteristics that make an experiment particularly pursuitworthy *qua* exploratory experiment? How do these characteristics differ from those that make an experiment pursuitworthy for theory testing?

### 3.3 What Is Experimental Pursuitworthiness?

To address such questions, we need an account of experimental pursuitworthiness. More specifically, the "unit of appraisal" (McMullin 1976) in pursuitworthiness assessments should not merely be theories, we should also look at experiments. This immediately raises the question of what an experiment is. The situation is complicated by the fact that the term ‘experiment’ is used in the sciences in a wide variety of senses. It can refer to specific interventions, to in-lab experimental setups but also to whole research programs such as those associated with big particle colliders. Questions of pursuitworthiness come up regarding all three senses. Moreover, pursuitworthiness questions can concern experimental methods, specific instruments, and specimens. In what follows, I will be interested in experiments as relatively stable facilities that allow scientists to perform a variety of interventions on their test objects.[19]

Next, it will be useful to consider in some more detail what exactly is meant by the ‘context of pursuit’ in experimentation. First, experimental pursuitworthiness should be distinguished from feasibility. Pursuitworthiness concerns whether an experiment should be performed. Feasibility concerns whether the experiment can be conducted at all. The pursuitworthiness of an experiment, thus, presupposes its feasibility. For example, the above-mentioned Solar-System-sized particle collider is simply unfeasible. Therefore, questions regarding its potential pursuitworthiness do not come up in the first place.[20]

Second, a natural place for issues of experimental pursuitworthiness is before or at the start of experimentation. Before scientists are ready to invest resources and time into a new project, they will want to assess the potential outcomes, the likelihood of achieving such outcomes, and the concrete costs associated with them. While feasibility concerns whether an experiment can at all be performed, questions of experimental pursuitworthiness are rather contrastive in nature. Which of the many possible experiments should be performed? Moreover, questions of pursuitworthiness do not just concern what experiments are to be performed but also how they should be performed, such that their epistemic outputs are maximized while their costs stay minimal.

Finally, issues of experimental pursuitworthiness also come up at later and more developed stages of experimentation, especially when scientists consider concluding experimentation. One reason to stop experimentation is the arrival at a stage where an experimental outcome is ready to be accepted or rejected because the gathered data makes the result statistically significant. What that amounts to differs from discipline to discipline and sometimes even from case to case (Fischer & Lossau, 2025). In particle physics, for example, the current criterion for accepting a discovery claim is a statistical significance of 5 sigma. 5 sigma corresponds to a p-value of 0.0000003. This p-value, in turn, represents the probability of a data pattern at least as extreme as that observed under the null hypothesis (the hypothesis that the novel phenomenon does not exist). In other words, this is a measure for how likely (or rather unlikely!) it is that the observed effect is merely a result of a statistical fluctuation. Once this threshold is reached, no more data is needed to accept the claim that a new particle was discovered, and thus no more experiment is needed to support that claim. Further experimentation can bring additional epistemic benefits such as more detailed characterizations of the newly discovered particle. The benefits related to the particle's discovery, however, are exhausted at that point.

[19] Some of the main lessons will extend to instances in which intervention in a strict sense is not possible, as in astronomy.
[20] It should be noted, however, that in practice the line between feasibility and pursuitworthiness is not always a sharp one. In what follows, I will argue that the pursuitworthiness of an experiment decreases with its costs. When such costs cross some threshold, questions of feasibility arise.

The opposite case is one in which no more experiment is to be carried out because past attempts at providing a statistically significant verdict regarding the hypothesis failed and it is deemed unlikely that further research will succeed. Again, this is the case when the costs of further experimentation are no longer justified by the epistemic benefits that one may expect. But in this case inquiry is concluded without a clear verdict regarding the hypothesis.

### 3.4 Criteria for Experimental Pursuitworthiness

What are the criteria or guidelines that should govern decisions about experimental pursuits? How do and should scientists make such decisions? While experimentation has not been an explicit concern in the philosophical literature on scientific promise and pursuitworthiness, there are, of course, resources to draw from. In what follows, we will see that insights from the philosophy of experimentation can be brought to bear on discussions of experimental pursuitworthiness.

The focus here lies on allocating scarce resources according to what novelty is most relevant. To conceptualize such decisions, I will use the economic framework introduced towards the end of chapter 2 as a structuring device: in general, an experiment's pursuitworthiness increases with the associated expected epistemic gain, and it decreases with the experiment's costs. The real challenge, of course, consists in spelling out what those gains and costs are and how they are to be determined in practice.

Consider expected epistemic gain. The expected epistemic gain depends on (1) the epistemic gains that are associated with the experiment's potential outcomes and (2) the conditional probability of achieving that outcome, given that the experiment is performed. I will focus here on the potential epistemic gains and later return to the likelihood of achieving those gains, in a separate section on uncertainty (section 3.5). The potential epistemic gains can be of various kinds, including (among other things) new and more data, and the discovery of new phenomena. Eventually such data and phenomena will be valuable in so far as they can be related to new or extant theories, models, and hypotheses. Epistemic gains can also be of an indirect character. They can consist in establishing new instrumentation, the development of novel methods and researcher training.

Costs are both financial and non-financial. Financial costs are the funds required for performing research, for example, for setting up and running a lab with required facilities and researchers. Cost assessment in experimentation is always highly context dependent. Costs will depend on the infrastructure that is already in place in the lab in which it is pursued. It depends on the local availability, for example, of lab technology, technical personnel and relevant specimens. Apart from financial costs considerations of pursuitworthiness should also reflect non-financial costs associated, for example, with moral hazards. Moreover, pursuit of an experiment can come with additional costs that derive from withholding a verdict regarding the hypothesis under consideration. This is especially the case in applied fields like regulatory science, where the decision to withhold a verdict may lead to delays in the availability of a new drug (Han, 2026).

The difficult part lies in giving guidelines for how to assess and weigh epistemic benefits and the costs associated with experimentation. In general, this is a difficult endeavor because epistemic gains are usually (1) hard to compare with each other, and (2) hard to compare with costs, especially in foundational research that is not exploited directly for economic purposes. These difficulties impose principled limitations on the very possibility of pursuitworthiness assessments in experimentation. Yet, despite these limitations we will see that some overarching considerations can still be made.

Towards the end of the previous chapter, I reviewed proposals to identify theoretical virtues as suitable indicators of theory promise. Douglas, for example, identified "scope, simplicity, and (potential) explanatory power" (2013, 800) as suitable candidates. Likewise, Duerr and Fischer (2025) propose to employ such virtues for assessing the epistemic gains and costs associated with theory development.

Here I will explore the prospects of extending that approach for assessing experimental pursuits. The idea is that, like theoretical pursuits are guided by theoretical virtues, experimental pursuits are guided by a distinct set of epistemic and pragmatic virtues that can be identified as experimental virtues.[21] In what follows, I will distinguish between experimental virtues that track an experiment's expected epistemic gains and virtues that guide the assessment of costs. In general, an experiment will be suited to yield epistemic gains if (or insofar as) its results clearly reflect a signal of a phenomenon of interest, its results are precise, and it facilitates direct observation. Moreover, the experiment benefits from having a broad sensitivity.

As a first virtue, consider a high signal-to-noise ratio. Any experimental signal is usually affected by a variety of background noise that results from the measurement instrument and other interfering processes. Signal clarity is what enables the experimentalist to generate insights about the phenomenon of interest. Therefore, an experiment will improve insofar as it succeeds in isolating the phenomenon of interest against any background processes. Alternatively, the backgrounds can be measured independently or modeled such that they can be subtracted from the recorded data (Boyd and Matthiessen 2024).

Second, consider precision. Precision is the "closeness of agreement between independent test results" (ISO, 2023, Section 3.12). Importantly, "precision depends only on the distribution of random errors and does not relate to the true value [...]" (ibid.). A measurement result can thus be very precise but highly inaccurate, that is, far from the true value. But in any case, the more precise the results of an experiment are, the more informative they are. Therefore, every increase in precision will also increase the epistemic gain associated with the experiment.

Third, experimental results or observations can be more or less direct. Generally speaking, the epistemic gain of direct results or observations is higher than that of indirect results or observations. To draw inferences from indirect observation we typically need to employ additional theoretical assumptions that are not required in instances of direct observation. And if these theoretical assumptions are false, then they undermine the evidential value of those inferences. The relevance of the direct-indirect distinction for considerations of experimental pursuitworthiness importantly depends on the epistemic status of the underlying theoretical assumptions. The epistemic gain of an indirect observation will not be threatened if theoretical assumptions are well-established. But the more speculative the assumptions are, the more they will inhibit potential epistemic gains.[22]

---

[21] The term "experimental virtues" has recently been coined by Peter Mättig and Michael Stöltzner (Mättig and Stöltzner 2025), who provide a list of criteria to assess experiments. The goal of the current discussion is more specific in that it is concerned with virtues that are related to considerations of pursuitworthiness. A more detailed argument for experimental virtues as guidelines for experimental pursuitworthiness is presented in Fischer (forthcoming).

[22] For a more detailed discussion of the direct-indirect distinction see Dudley Shapere's (1982) account of direct observations of the centre of the Sun. More recently, the epistemic role of the distinction has been

Finally, broad sensitivity means that the experiment can distinguish between a large number and variety of conditions. Broadness of sensitivity promotes the expected epistemic gains of an experiment in at least three distinct ways. First, it enhances the experiment's ability to characterize the phenomenon in a variety of ways (Boyd and Matthiessen 2024). Consider once more the example of Ampère's exploratory experiments with the astatic needle. Ampère's ability to vary the relative position of the needle allowed him to characterize the wire's magnetic field in a comprehensive way (Steinle 1997). Second, broadness of sensitivity increases an experiment's ability to produce unexpected results. For example, consider the surprise discovery of pulsars at the Mullard Radio Astronomy Observatory in 1967 (Hewish et al., 1968). Pulsars appear as pulsating sources of electromagnetic radiation. The surprise discovery was facilitated by high time resolution achieved through short integration time recorders, which were unusual then in radio telescopes (Penny, 2013).Third, broadness of sensitivity enables cross checking (Mättig and Stöltzner 2025). If an experiment is sensitive to a variety of phenomena one can increase one's confidence in the apparatus by calibrating it on known data, before it is employed to produce novel data.

This list of virtues can easily be extended. Additional criteria are the *absence of* or the *ability to control systematic errors* (Staley, 2020), an experiment's *technological novelty* (Laymon and Franklin 2024), its interactivity (Murphy et al., forthcoming), and its *replicability*. Moreover, the current set of virtues is heavily inspired by examples from physics. Additional, (although certainly not entirely unrelated) criteria can be found in other disciplines. Looking at experimentation in biochemistry Vallejos-Baccellieri (2026) identifies an experiment's capacity (1) for generating and expanding data domains, (2) for enabling fine-grained causal control, and (3) for reducing theoretical dependence on experimental access to phenomena as relevant experimental virtues. Moreover, Nevia Dolcini and colleagues identify methodological robustness, statistical power, and the interpretability of results as important criteria in neurolinguistics (Dolcini et al., 2025).

So far, we have been concerned with virtues that indicate an experiment's expected epistemic gain. Similar overarching virtues can be identified as relevant for estimating experimental costs. Examples for such virtues are simplicity and continuity with extant practices. Simplicity, of course, relates to experimental costs in a very direct way: simple experiments are cheaper to set up and run.

For example, consider precision measurements of the electron magnetic moment $g$. For physicists $g$ is of interest because the Standard Model of particle physics makes extremely precise discussed in the philosophy of particle physics (A. Franklin, 2017) and gravitational wave astronomy (Ahmed, 2025; Doboszewski & Lehmkuhl, 2023; Elder, 2025).

predictions for *g* and those predictions can be tested by so-called Penning traps (Brown & Gabrielse, 1986; Hanneke et al., 2008; Odom et al., 2006).[23] Nick Huggett and Mike Schneider (2025) contrast the simplicity and small size of a Penning trap experiment with the vast dimensions of the Large Hadron Collider: "the equipment used was the size of a small car, not of a small country (and run by a team of the size of a carpool, not the size of the population of a UN recognized microstate)" (6). While measurement of the electron magnetic moment requires high precision and elaborate elimination of interfering factors, it is—compared to the Large Hadron Collider—still relatively simple in a technological and organizational sense. An individual scientist can still maintain an overview of all the parts of the Penning trap, while experimental capabilities need to be distributed over a whole research community in the case of the Large Hadron Collider (ibid., 10). Recent empirical research has shown that the technological complexity of such a large machine also leads to new levels of organizational complexity. For example, a detector like the ATLAS (A Toroidal LHC ApparatuS) detector at the Large Hadron Collider requires an organizational approach to risk management, implemented in dedicated technical review procedures (van Panhuys & Jadreškić, 2026).

The (relative) technological simplicity of a Penning trap should be distinguished from inferential simplicity that has been of recent interest in discussions of the aesthetics of experiment (e.g., Ivanova, 2021, 2023; Ivanova et al., 2024). For example, the Meselson-Stahl experiment has been identified as a particularly simple experiment because of the direct inferences it licenses regarding the replication mechanism of DNA. Unlike technological simplicity, this form of inferential simplicity is better described as a benefit-indicating virtue rather than a cost-indicating virtue. This is because it promotes the experimentalist's understanding of the research object and thus increases the epistemic value of the experiment's outcomes.

Another overarching virtue that will help control an experiment's costs is its continuity with extant experiments and experimental practices. Insofar as an experiment continues existing traditions it allows the transfer of skills and of material and organizational resources. One might think that such continuity stands in direct opposition to what I have identified as the virtue of technological novelty above. That does not need to be the case, however. An experiment can be novel in important regards while still making use of extant resources and practices.

For example, consider the Future Circular Collider, which is currently considered as a follow-up experiment to the Large Hadron Collider. Stephen Myers, engineer in high-energy physics, has argued that the Future Circular Collider is promising because it continues a successful research tradition of circular colliders, while enabling the exploration of particle collisions of unprecedented energy (2021). According to Myers, continuity is particularly important because there were instances in which a lack of such continuity inhibited or even prevented success. First, consider the Stanford Linear Collider (SLC), the first linear electron-positron collider. The SLC started operation in 1989 and stood in direct competition with the Large Electron Positron Collider (LEP), a circular collider at CERN (European Organization for Nuclear Research). Myers argues that

> since the SLC was the first ever linear collider, the luminosity learning curve was shallow and tortuous, whereas LEP rapidly reached its design luminosity using the well-tried and tested synchrotron collider design and the vast amount of experience gained in previous similar colliders. Consequently, LEP very quickly outperformed the SLC in the production of high luminosity. In my opinion, the lesson from this story is that the technology leap was too big for a new type of collider which needed to produce data for physics on the same timescale as a more traditional albeit much larger collider that was LEP. (ibid., 4)

[23] See Koberinski and Smeenk (2020) for a more detailed characterization of the experiment aimed at philosophers.

Myer's second example is the failure of the Superconducting Super Collider (SSC) that was supposed to be built in Waxahachie, Texas. Construction of the SSC started in 1991 and continued until 1993, when the project was shut down due to an explosion in projected costs (Riordan et al., 2018). Myers identifies several reasons for the project's failure, including poor planning and disagreement about it in the scientific community. But a central reason for the failure, according to Myers, was that "Waxahachie was a 'green field' site without existing accelerators and poor infrastructure" (7). This is a problem because the "development of an accelerator complex like that of Fermilab, Stanford, Brookhaven and CERN takes decades" (ibid.). Moreover, according to Myers, "if the Fermilab site had been chosen for the SSC, the project would have been a big success for US particle physics" (ibid., 6). Whether Myers's claims, especially that latter counterfactual claim about SSC's success, hold up to scrutiny need not concern us here. What matters for our purposes is that they illustrate how continuity with extant experiments has been considered as an important factor influencing a project's costs and feasibility.

Any assessment of experimental costs will depend on many other factors besides simplicity and continuity. Further examples are *modularity* and *speed*. An experimental facility's modularity allows upgrading and cost-effective repurposing of parts of the experiment. Experimental speed helps control costs by minimizing time spent on the experiment before intended results are achieved. Moreover, there are further discipline-specific virtues that affect cost assessment, such as participant accessibility in experimentation involving human test subjects (Dolcini et al., 2025).

The overall idea here is to identify experimental virtues as promise reasons, that is, as criteria that indicate whether experimental efforts will be rewarded adequately or not. But do these virtues represent properly epistemic reasons for pursuit? Benefit-indicating virtues are unproblematic in that regard. Signal clarity, precision, directness and broadness of sensitivity all promote the improvement of knowledge in a direct way. Cost-indicating virtues such as simplicity and continuity, however, seem to have a more pragmatic nature. Nevertheless, these virtues can be characterized as epistemic in the sense discussed in section 2.3. Insofar as they help generating maximal epistemic output at minimal costs they are "in function of epistemic goals" (Šešelja et al. 2012, 67). Moreover, unlike the reasons in Fleisher's sandwich-shop scenario, they cannot be dismissed as promoting pursuitworthiness only marginally or in an overly specific sense.

How do these virtues affect decision making in experimentation? Mättig and Stöltzner argue that just like theoretical virtues such as empirical adequacy, scope and simplicity govern theory choice, experimental virtues "guide experimentalists' choices without being rationally compelling" (2025, 8). Moreover, like Kuhnian values of theory choice they have an overarching scope, they "were developed by experimental practice and persist across major changes in measurement devices and methods" (ibid.). Another feature that theoretical and experimental virtues have in common is that they are affected by what Kuhn (1977) calls "conflict and equivocation" (362). When two agents subscribe to the same set of virtues, they may still disagree about what follows from those virtues. This is because agents may disagree about how virtues are to be applied in concrete instances. For example, if two agents agree that an experiment should be as simple as possible, there can still be considerable disagreement about whether that refers to technological or inferential simplicity.

Like theoretical virtues, experimental virtues can also stand in tense trade-off relationships. For example, usually an increase in signal-to-noise ratio or precision requires eliminating the influence of interfering factors. This usually makes an experiment more complex and more expensive. Moreover, precision and broadness of sensitivity can be conflicting virtues. The muon $g$-2 experiment at Fermilab, for example, determined the anomalous magnetic dipole moment of the muon to unprecedented precision (Aguillard et al., 2025). But the experiment is not broadly sensitive, as determining that constant is its sole purpose.

How virtues are to be traded off against each other depends among other things on the purpose of experimentation. In hypothesis testing the experiment should warrant a clear verdict on the hypothesis at reasonable costs. This is achieved if the experiment provides a clear signal with high precision while not being overly complex and making use of extant resources and practices. In exploratory experimentation, broadness of sensitivity is particularly important such that a wide range of parameters can be covered, and the experiment should be modular such that it can be modified at minimal costs to adapt to novel insights that are gained during inquiry. In exploratory experimentation signal-to-noise ratio and precision will certainly also play a role, but they may be traded more readily against the experiment's broadness of sensitivity and interactivity than in hypothesis-testing.

Let me conclude this discussion with a note on the comparisons that can be made in the presented framework. When experiments compete for addressing the same research questions, the above-discussed virtues give some overarching guidance in decision making as to what experimental procedures are to be prioritized. However, experiments often compete for the same resources even if they have different epistemic goals. If a Penning trap competes for the same resources as a large particle collider one cannot just point to the Penning trap's simplicity to argue that it is more pursuitworthy. Instead, one would need to provide a more general assessment of the epistemic gains to be expected from probing high energy levels as opposed to precision measurements at low energy levels (Koberinski, 2026). Such comparisons become increasingly difficult the more the research questions differ. If two experiments from widely different fields compete for the same resources, there may simply be no meaningful comparison of epistemic pursuitworthiness.[24]

### 3.5 Uncertainty and the Pursuitworthiness of Experiments

So far, I have discussed the role of potential epistemic gains and costs in assessments of experimental pursuitworthiness. Here I will address additional challenges that arise from uncertainty in assessing an experiment's pursuitworthiness. The pursuitworthiness of an experiment depends on how valuable we expect its epistemic output to be. However, usually we have only limited knowledge of the outcomes that are to be expected and how highly we will value them. More specifically, there are three kinds of uncertainty.

**Outcome uncertainty** is an uncertainty that is associated with what the results of an experiment will be. Typically, it is not known in advance what the outcome of an experiment will be.

[24] Note that this is not just a problem for experimental virtues. For the same reasons we do not employ theoretical virtues to compare Newtonian Mechanics with Evolutionary Biology. The difference is that there is no sense in which Newtonian Mechanics and Evolutionary Biology compete, whereas physics experiments can compete for the same research funds as biology experiments.

**Modal uncertainty** concerns what results of an experiment are possible in the first place. An experiment can have results that are unaccounted for in theorizing and experimental planning.

**Value uncertainty** concerns the epistemic gain associated with the experimental result. What looks like a potentially valuable result in advance may turn out to be of little epistemic value once it is obtained as the result of experimentation. What looks like a result that is of little interest, may turn out to represent an astonishing empirical advancement.

While each of these kinds of uncertainty poses challenges to reliable ex-ante assessments of experimental pursuitworthiness, they do not make such assessments impossible. Consider *outcome uncertainty*, which is associated with our lack of knowledge of what the experiment's result will be. Ignoring for a moment modal and value uncertainty we model the overall expected epistemic gain associated with experiment *E* as follows:

$$EEG(E) = \sum p(O_i) \cdot EG(O_i).$$

Here $EG(O_i)$ represents the epistemic gain associated with each of the experiment's potential outcomes $O_i$, that are here assumed to be mutually exclusive and jointly exhaustive of the realm of possibility. Each of these epistemic gains contributes to the overall expected epistemic gain weighted by the probability $p(O_i)$. This probability, in turn, depends on how likely it is taken to be that the corresponding experimental result $O_i$ is achieved.

A few notes on this probability. First, since we are concerned with ex-ante assessments of pursuitworthiness, these probabilities should reflect the subjective state of belief of the scientist who is evaluating the experiment's pursuitworthiness. It may happen that scientists disagree about these probabilities and thus come to conflicting assessments of an experiment's pursuitworthiness.

Second, the probability is a conditional probability—conditional on the experiment being performed. This introduces further challenges insofar as there may be disagreement about what performing the experiment amounts to.

Third, the probability involves a convolution of two factors. First, the assumed likelihood of the existence of result $O_i$ and, second, the sensitivity of the experiment required to establish the result. If the outcome likely exists but the experiment is bad at picking up the signal that would indicate that outcome, then the experiment's overall expected epistemic gain will be low. Likewise, it will be low if the experiment is sensitive to those potential outcomes that are very unlikely to exist. Ideally, the experiment will be sensitive to all potential outcomes. If that is not possible, then the experiment should maximize sensitivity regarding those potential outcomes that are particularly likely.

So-called 'crucial experiments' are good examples of cases where this formalization can be applied. The concept of crucial experiment goes back all the way to Bacon's epistemology of experimentation and describes the situation where an experiment addresses two mutually exclusive and jointly exhaustive hypotheses. Falsifying one of the two hypotheses leads to confirmation of the other hypothesis.

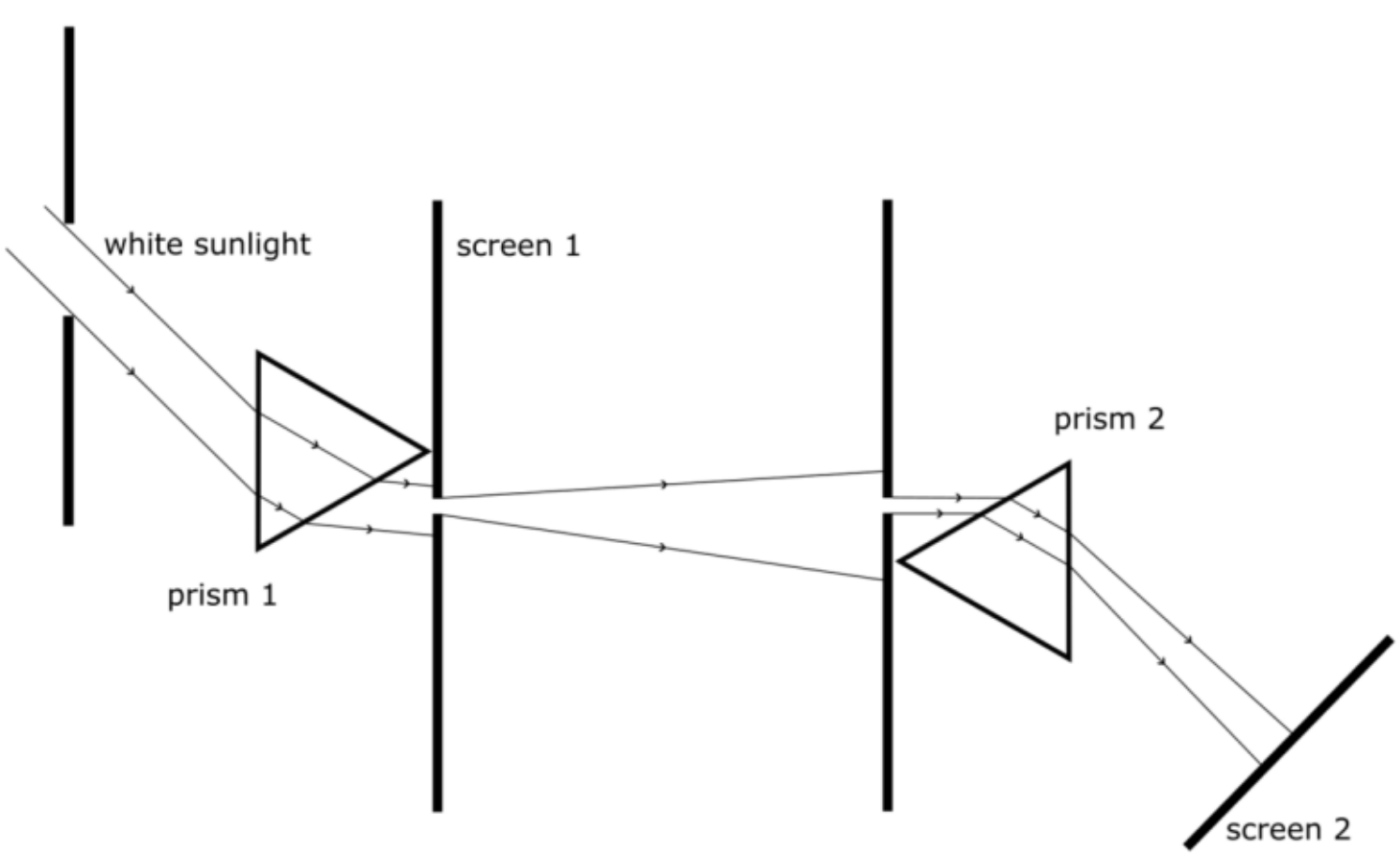

*Figure 2: Schematic representation of Newton's crucial experiment on the heterogeneity of white sunlight.*

As an illustration consider the famous example of Newton's prism experiment (Currie and Levy 2019). At stake were two mutually exclusive hypotheses regarding the nature of white sunlight. The prevailing hypothesis was that white sunlight is homogenous, meaning that it is composed of rays with identical properties. Today we know that white sunlight is heterogeneous, meaning that it is composed of rays of different colors. Newton showed this by isolating a beam of sunlight and sending it through two prisms (see figure 2). Sending light through the first prism created a full color spectrum on a screen behind that prism. That screen had an aperture which let part of that spectrum through to be transmitted to the second prism placed behind that screen. The light going through that second prism was projected on another screen located behind the second prism. By manipulating the angle of the first prism Newton could select what part of the spectrum was transmitted through to the second screen. Newton reasoned that if light is homogeneous, then the position of the light beam on the far screen does not depend on what part of the beam is filtered. Thus, altering the angle of the first prism would have no effect on the position of the transmitted light on the second screen. But changing the angle of the first prism *did* influence the position of the beam on the second screen, meaning that refraction did depend on the part of the spectrum that was transmitted. This, in turn, implied that light is heterogeneous.

Newton's experiment here is modeled as a case in which there are two mutually exclusive and jointly exhaustive hypotheses or outcomes. Moreover, the experiment is assumed to be clearly sensitive to distinguishing between the two hypotheses. Insofar as this is an adequate model of Newton's experiment, the summed probability of the two considered outcomes is 1, meaning that the experiment guarantees an epistemic output if it is performed. In other words, one reason why ideal instances of crucial experimentation are epistemically favorable is that they allow a special kind of control over experimental uncertainty: no matter what the outcome of the experiment is, the experiment produces a clear verdict on the two competing hypotheses.

Of course, the ideal conditions of a crucial experiment are rarely if ever instantiated. More commonly, one will have only limited understanding of the various potential outcomes, their likelihoods, and the relation between the outcomes. Moreover, experimentation usually is not equally sensitive to all potential outcomes, and it rests on a wide range of auxiliary hypotheses on which one may err. But even if hypotheses are not mutually exclusive and jointly exhaustive, and even if experimentation depends on additional assumptions, experimenters can devise strategies to maximize the experiment's epistemic payoff, by controlling outcome uncertainty. More specifically, experimenters can work towards a no-lose situation, that is, a situation in which each of the experiment's foreseeable outcomes represents a significant epistemic advancement.

An example of such a no-lose situation is the search for the Higgs boson (for details see Fischer (2026a)). The discovery of the Higgs boson in 2012 was an important result for particle physics. The corresponding Higgs mechanism had been predicted already in the 1960s (Englert & Brout, 1964; Higgs, 1964). Yet there was no overwhelming agreement that the Standard Model Higgs boson would be discovered at the Large Hadron Collider (Mättig & Stöltzner, 2019). There were, however, strong theoretical arguments to the effect that the Large Hadron Collider had to find something—even if the Higgs could not be observed. More specifically, if the Higgs had not shown up in the energy range probed by the Large Hadron Collider, then some other novel phenomenon would have been required to save unitarity, a core principle of quantum mechanics. In the words of physicist Sally Dawson (2022) who remembers the situation of the field in the 1980s:

> "*The LHC [Large Hadron Collider] and the SSC [Superconducting Super Collider] had to find something*: If the Higgs boson weren't found, the thinking went, then there would be supersymmetry, technicolor, or something new at the energy scales accessible to the LHC and SSC that would restore perturbative unitarity. It is impossible to overemphasize the impact of this reasoning as it triggered serious thinking about how the Higgs boson could be observed at a hadron collider."

The example of the Higgs boson is one where experimentation is guided by clearly formulated theoretical principles, something that is not always reproducible in other fields, or even within particle physics. However, the overarching lesson here still is that it is a matter of well-prepared experimentation to make sure that one can learn something from the experiment even if the favored result is not obtained, or if there is no favored result in the first place. What matters for this is a clear expectation of the possible results of the experiment, their likelihoods and their associated epistemic gains. This overall lesson extends to other forms of experimentation, for example, exploratory experimentation. Even if there are no strictly foreseeable outcomes in exploratory experimentation most instances of exploration are not simply shots in the dark: these efforts are motivated and guided by expectations as to what can be found in the parameter regime that is being probed.

The second kind of uncertainty, modal uncertainty, concerns what are even possible outcomes of the experiment. In the formal framework this amounts to the question of what $O_i$ are to be included in the formula for the expected epistemic gain. Consider surprising discoveries such as the discovery of the cosmic microwave background by Penzias and Wilson or the discovery of x-rays by Röntgen. These results did not show up in any ex-ante considerations of the pursuitworthiness of the respective experimental setups. Consequently, it seems that modal uncertainty is much harder to control. Yet generally such surprising discoveries can be facilitated by experimental designs that have broad sensitivity and allow the experimenter to vary many parameters.[25]

More difficulties arise in attempts to control value uncertainty, the kind of uncertainty that is associated with changes in the epistemic value that is assigned to an experiment's outcome. Suppose an experiment is executed successfully, and the predicted result is observed and consistent with expectations. Then the epistemic value of the result can still deviate from initial expectations, because our valuing of that result has changed in the meantime.

Such changes can come up in two distinct ways. First, the evaluation may change through additional insights into the result itself. For example, the discovery of the Higgs boson was the result of a successfully executed experiment, it was to a certain degree expected, and the discovery itself represents a major achievement as it confirms the Standard Model of particle physics. Yet, it is, in a sense also disappointing, since it did not indicate physics beyond the Standard Model, as many physicists had hoped. The result of an experiment may also be more valuable than expected. When Penzias and Wilson were calibrating their radio astronomic antenna in 1965, they failed to eliminate or explain all background noise in their measurement device. What initially looked like a nuisance potentially related to pigeons living in the horn antenna, eventually turned out to be cosmic background radiation that is highly valuable for cosmology.

[25] Surprise discoveries also require conceptual openness of the research program. For an example, in which conceptual rigor can inhibit progress see DiMarco's (2025) discussion of sex as a biological variable in exploratory experimentation.

Second, the standards for what is to count as an epistemic advancement may change through the experiment. In such cases the result may be exactly as expected, but the assessment of the result as valuable has changed because the relevant standards have changed. The situation is to some degree analogous to transformative experiences (Paul, 2014). Decisions that fundamentally change one's life—such as becoming a parent—can also affect the criteria that one employs to evaluate those decisions.

Kuhnian paradigm shifts are examples of such transformations. According to Kuhn (1962), a paradigm does not only define the empirical problems that a scientific community should address but also the criteria that are employed to assess answers to those problems. If the paradigm changes, then those evaluation criteria may change as well. For example, Galileo's telescope observations were not considered pursuitworthy in the paradigm of Aristotelian physics. Those observations, however, helped overthrow that Aristotelian paradigm in favor of Galilean physics, according to which those observations are pursuitworthy.

According to L. A. Paul, transformative experiences pose a challenge to decision making. If one relies on one's initial set of criteria, one risks making a decision that does not align with one's altered preferences. Attempts to base one's decision on the altered preferences, however, are limited by the access one has to the future preferences. Analogously, a scientist's decision to perform an experiment may be evaluated differently if the very standards by which the experiment is to be evaluated are affected by the experimental result. If one relies on one's initial assessment of epistemic gains, one risks making a decision that does not align with the altered evaluation criteria. Attempts to employ the altered evaluation criteria, however, are limited if one does not know those results in advance.

Scientists' options to control value uncertainty are limited. As a remedy, Szymon Miłkoś (2026) suggests evaluating not the promise of the ideas that are being researched but the promise of the pursuit itself. This is facilitated, according to Miłkoś, by a "Foundational Norm of Transformative Pursuitworthiness": scientists should document and make transparent their private assessments of what they consider pursuitworthy. Thus, scientists' personal hunches can be made subject to community assessment and may transform into public reasons. Further remedy may be obtained from the study of future scenarios and by explicit mapping of science's possible futures (Virmajoki, 2023). It should be noted, though, that such methods are limited by our epistemic access to any such scenarios.

Let us take stock. Extant discussions of pursuitworthiness have to a large degree focused on the promise of theories and theoretical parts of scientific practice. This chapter has addressed additional questions that arise regarding the pursuitworthiness of experiments. The starting point of the discussion here was that there are typically many more possible experiments than can be performed. Resource scarcity, thus, calls for a cost-benefit analysis of experimental pursuits. Integrating key insights from the philosophy of experimentation I have discussed a series of experimental virtues that guide assessments of costs and epistemic benefits. Moreover, we have seen that assessments of pursuitworthiness are affected by three kinds of uncertainty: uncertainty related to the experiment's outcome (outcome uncertainty), uncertainty related to what even the possible outcomes of an experiment are (modal uncertainty), and uncertainty regarding the evaluation of the epistemic gain associated with such outcomes (value uncertainty).

I have employed the economic model here as structuring device to shed some light on experimental pursuitworthiness. However, that approach itself has its limitations, as some of the discussion has already highlighted. Further limitations arise because the framework so far has construed epistemic gains as associated with concrete experimental outcomes. But experimentation can lead to many kinds of epistemic gains that are not necessarily related to concrete outcomes. For example, Vijay and Arrabito (2026) criticize the current model for not representing long-term epistemic payoffs that are not related to specific outputs. Vijay and Arrabito argue that this does not capture the value of "infrastructure projects like genome databases [that] may only realize their epistemic value years later through meta-analyses or unforeseen applications" (2026, 17). Thus, the discussion of experimental pursuitworthiness put forward in this chapter is far from conclusive and will hopefully inspire further thought.

**Chapter 4. The Future of Pursuitworthiness**

Our starting point was an apparent dilemma: trying to choose the most fruitful line of research is like putting 'the cart before the horse' because we cannot know the outcomes of a line of research before pursuing it. Moreover, it seems scientists regularly need to go against good reasons when they decide to research new theories and methods because they could have worked on better-established competitor theories instead. The reviewed discussion of pursuitworthiness has yielded answers to both these concerns. While uncertainty remains, choosing one's line of research is not entirely like putting the cart before the horse, because there are indicators of scientific promise that give important overarching guidance. Moreover, such indicators can help justify scientists' choices to invest their efforts into theories that are currently less acceptable than their strongest alternatives.

Large parts of the extant philosophical literature on pursuitworthiness have focused on theory pursuit. The current Element has shown that this overly narrow focus on theory pursuit is problematic for two reasons. First, theoretical pursuitworthiness in important ways depends on experimental pursuitworthiness. Second, experimentation raises novel issues for thinking about scientific promise. To address such issues, the Element has provided an account of experimental pursuitworthiness that rests on evaluating experimental virtues—overarching features of experiments that can be employed as proxies for the expected epistemic benefits and costs of experimentation. Moreover, the Element has developed a novel taxonomy of uncertainties that are encountered in assessments of experimental pursuitworthiness.

A brief introduction to a topic as large as that of scientific promise cannot do much more than scratch the surface. There are many places one could dig deeper. The current Element has approached the issue of scientific promise on a general epistemological level. Deeper insights will be gained by looking at pursuitworthiness through the lens of case studies of concrete pursuits. Such case studies would show local determinants of pursuitworthiness that complement the general indices of theory promise that have been discussed here. Theory building in physics, for example, is often guided by overarching physical principles such as the (generalized) correspondence principle (Dardashti et al., 2025; Fischer, 2024a). These principles fall short of being general indicators of pursuitworthiness because they do not apply to all cases of theory development. Yet, on the local level they clearly tell physicists what kind of theories they should pursue and what theories can be rejected right away. A discussion of such local factors would also be required for generating more specific recommendations of what projects are pursuitworthy and what projects are not.

The Element has deliberately sidelined a few issues that would have deserved coverage in a larger volume. First, the Element has not discussed questions of non-epistemic or practical pursuitworthiness. Demarcation is particularly difficult in cases of fields that respond to direct societal needs such as biomedical research (Fisher, 2023), sustainability research (Stojanovic, 2023), and solar geo-engineering (O'Loughlin & Visioni, 2026). Additional issues of pursuitworthiness arise in emergency situations such as a global pandemic where a quick response from science is needed. The pursuitworthiness of, e.g., epidemiological modeling or vaccination research is evident in such situations. Yet, how does an emergency justify deviations from routine procedures such that science remains pursuitworthy (Stegenga, forthcoming)? The issue of pursuitworthiness also has obvious connections to the politics of science, issues of science planning (Baker, 2022), and questions of science funding (Shaw, 2021). To what degree should decisions about scientific pursuits be left to scientists? What are suitable funding mechanisms to ensure that the most pursuitworthy research gets done? Finally, the book has also been limited to philosophical discussions of scientific pursuitworthiness. The issue has of course also been taken up in other disciplines such as science economics (e.g., Stephan, 2012) and science and technology studies (STS), even if not under the label of 'pursuitworthiness.' Existing discussions in STS of what makes a scientific problem do-able (Fujimura, 1987; Sorgner, 2022), for example, could easily be brought to bear on the more pragmatic considerations of experimental pursuitworthiness discussed in chapter 3.

The topic of scientific promise has been an exceptionally fruitful one for philosophy of science and it will continue to be so in the future. Let me conclude by highlighting just three exemplary lines of research that will likely attract philosophers' attention in the years to come.

First, as artificial intelligence is currently revolutionizing the ways science works it will also affect considerations about the pursuitworthiness of theories and experimental research programs. In many areas of science, the use of deep learning methods impacts pursuits by vastly decreasing the costs of research and increasing epistemic output. For example, the deep-learning-based protein-structure prediction system AlphaFold (Jumper et al., 2021) generates highly accurate predictions of the three-dimensional structure of proteins at an unprecedented speed. This novel tool enables new lines of research. It also affects how scientists evaluate pursuits because the tool's predictions "provide plausible hypotheses that can suggest mechanisms of action and allow designing of experiments with specific expected outcomes" (see also Curtis-Trudel et al., 2025 for further discussion; Terwilliger et al., 2024, p. 115).

Second, the pursuitworthiness of Big Science will be of major concern. Currently planning is underway for particle colliders such as the Future Circular Collider (FCC) at CERN. The FCC would be placed in a tunnel of 91 km in circumference with estimated costs of at least $30 billion and a projected timeline stretching over up to 70 years (Castelvecchi, 2025). Certainly, the pursuit of large projects such as those at CERN is not just a scientific matter but in large part one of international science politics. Yet even if the success conditions of such a project are to a large degree determined by political factors, there is an underlying question of what makes these projects epistemically pursuitworthy—compared, for example, with more small-scale projects. Additional issues of pursuitworthiness arise in other fields of Big Science. Looking at examples like the Human Genome Project, Vijay and Arrabito (2026) show that big biology faces specific challenges because it needs to "navigate irreducible complexity, decentralized collaboration, and evolving societal expectations" (1). For example, whereas particle physics experiments are performed at centralized facilities such as CERN, big biology projects need to integrate experimental results that are spread locally.

Third, the pursuitworthiness of experimentation is an exciting field for further studies. A central epistemological benefit of the notion of pursuitworthiness and associated discussions of the "context of pursuit" has been an improved understanding of the practices of theory development. Likewise, the notion of experimental pursuitworthiness can be employed to better understand concrete experimental practices that already exist. Consider exploratory data analysis (EDA). EDA is a set of techniques to draw preliminary conclusions from large data sets, including especially graphical methods (Tukey, 1977). Exploratory data analysis has a somewhat mixed reputation if understood as a kind of confirmatory practice: "exploratory data analysis might be what happens when you fail to specify hypotheses in advance, or test multiple hypotheses without adequate correction, or choose a lenient alpha for significance testing, or exploit freedom in parameter choice, or whatever" (Klein, 2024, p. 1114). However, EDA according to Klein (2024) is much better understood as a method for determining the "expected value of experimentation" (1109). Instead of confirming a hypothesis its main function is to direct scientists to the experiments that are worth performing. For short: it helps establish an experiment's pursuitworthiness.

Further examples of existing experimental practices for addressing questions of pursuitworthiness are pilot studies and feasibility studies. Consider pilot studies. Before conducting a full-scale experiment, scientists often devise pilot studies. The pilot study employs methods and procedures that are similar to those of the parent study. The goal is to test the methods and procedures and to anticipate problems that may arise in the conduct of the parent study. Thus, the pilot study helps estimate and maximize the expected epistemic benefits of the parent study, and it informs associated considerations of costs and feasibility. Moreover, large experimental programs such as the above-mentioned Future Circular Collider are being prepared by feasibility studies (see, e.g., Benedikt et al., 2025). Such studies, among other things, make a case for the pursuitworthiness of the planned research by highlighting the scientific opportunities it opens up and by surveying the technologies that will enable pursuit. Thus, the topic of scientific promise and associated debates of pursuitworthiness will keep philosophers of science busy for many years to come.